\documentclass[twocolumn]{aastex631}
\usepackage{graphicx}
\usepackage{xcolor}
\usepackage{xspace}
\usepackage{amsmath}
\usepackage{tikz}

\newcommand{\kms}{km s$^{-1}$\xspace}
\def\q#1{`#1'}

\newcommand{\app}{$\sim$\xspace}
\newcommand{\msun}{M$_{\odot}$\xspace}
\newcommand{\cms}{cm$^{-2}$\xspace}
\newcommand{\cmq}{cm$^{-3}$\xspace}
\newcommand{\asec}{$^{\prime}$$^{\prime}$\xspace}
\usepackage{calc}
\newlength{\availableHeight}
\newlength{\figureHeight}
\begin{document}

\title[3-D CMZ III]{3-D CMZ III: Constraining the 3-D structure of the Central Molecular Zone via molecular line emission and absorption}

\author[0000-0001-7330-8856]{Daniel~L.~Walker}\thanks{E-mail: \texttt{daniel.walker.astro@gmail.com}}
\affiliation{UK ALMA Regional Centre Node, Jodrell Bank Centre for Astrophysics, The University of Manchester, Manchester M13 9PL, UK \\}
\affiliation{Department of Physics, University of Connecticut, 196A Auditorium Road, Storrs, CT 06269 USA \\}

\author[0000-0002-6073-9320]{Cara Battersby}
\affiliation{Department of Physics, University of Connecticut, 196A Auditorium Road, Storrs, CT 06269 USA \\}
\affiliation{Center for Astrophysics $|$ Harvard \& Smithsonian, MS-78, 60 Garden St., Cambridge, MA 02138 USA}

\author[0000-0002-5776-9473]{Dani Lipman}
\affiliation{Department of Physics, University of Connecticut, 196A Auditorium Road, Storrs, CT 06269 USA \\}

\author[0000-0001-6113-6241]{Mattia C.\ Sormani}
\affiliation{Universit{\`a} dell’Insubria, via Valleggio 11, 22100 Como, Italy}

\author[0000-0001-6431-9633]{Adam Ginsburg}
\affiliation{University of Florida Department of Astronomy, Bryant Space Science Center, Gainesville, FL, 32611, USA}

\author[0000-0001-6708-1317]{Simon C. O. Glover}
\affiliation{Universit\"{a}t Heidelberg, Zentrum f\"{u}r Astronomie, Institut f\"{u}r Theoretische Astrophysik, Albert-Ueberle-Str.\ 2, 69120 Heidelberg, Germany}

\author[0000-0001-9656-7682]{Jonathan~D.~Henshaw}
\affiliation{Astrophysics Research Institute, Liverpool John Moores University, 146 Brownlow Hill, Liverpool L3 5RF, UK}
\affiliation{Max-Planck-Institut f\"ur Astronomie, K\"onigstuhl 17, D-69117 Heidelberg, Germany}

\author[0000-0001-6353-0170]{Steven N.~Longmore}
\affiliation{Astrophysics Research Institute, Liverpool John Moores University, 146 Brownlow Hill, Liverpool L3 5RF, UK}

\author[0000-0002-0560-3172]{Ralf S.\ Klessen}
\affiliation{Universit\"{a}t Heidelberg, Zentrum f\"{u}r Astronomie, Institut f\"{u}r Theoretische Astrophysik, Albert-Ueberle-Str.\ 2, 69120 Heidelberg, Germany}
\affiliation{Universit\"{a}t Heidelberg, Interdisziplin\"{a}res Zentrum f\"{u}r Wissenschaftliches Rechnen, Im Neuenheimer Feld 225, 69120 Heidelberg, Germany}
\affiliation{Center for Astrophysics $|$ Harvard \& Smithsonian, MS-78, 60 Garden St., Cambridge, MA 02138 USA}
\affiliation{Radcliffe Institute for Advanced Studies at Harvard University, 10 Garden Street, Cambridge, MA 02138, USA}

\author[0000-0003-4140-5138]{Katharina Immer}
\affiliation{European Southern Observatory (ESO), Karl-Schwarzschild-Straße 2, 85748 Garching, Germany}

\author[0009-0005-9578-2192]{Danya Alboslani}
\affiliation{Department of Physics, University of Connecticut, 196A Auditorium Road, Storrs, CT 06269 USA \\}

\author[0000-0001-8135-6612]{John Bally}
\affiliation{CASA, University of Colorado, 389-UCB, Boulder, CO 80309}

\author[0000-0003-0410-4504]{Ashley Barnes}
\affiliation{European Southern Observatory (ESO), Karl-Schwarzschild-Straße 2, 85748 Garching, Germany}

\author[0000-0003-0946-4365]{H Perry Hatchfield}
\affiliation{{Jet Propulsion Laboratory, California Institute of Technology, 4800 Oak Grove Drive, Pasadena, CA, 91109, USA}}
\affiliation{Department of Physics, University of Connecticut, 196A Auditorium Road, Storrs, CT 06269 USA \\}

\author[0000-0001-8782-1992]{Elisabeth A.~C.~Mills}
\affiliation{Department of Physics and Astronomy, University of Kansas, 1251 Wescoe Hall Drive, Lawrence, KS 66045, USA}

\author[0000-0002-0820-1814]{Rowan Smith}
\affiliation{Scottish Universities Physics Alliance (SUPA), School of Physics and Astronomy, University of St Andrews, North Haugh, St Andrews KY16 9SS, UK}
\affiliation{Jodrell Bank Centre for Astrophysics, Department of Physics and Astronomy, University of Manchester, Oxford Road, Manchester M13 9PL, UK}

\author[0000-0002-9483-7164]{Robin G. Tress}
\affiliation{Institute of Physics, Laboratory for Galaxy Evolution and Spectral Modelling, EPFL, Observatoire de Sauverny, Chemin Pegasi 51, 1290 Versoix, Switzerland}

\author[0000-0003-2384-6589]{Qizhou Zhang}
\affiliation{Center for Astrophysics $|$ Harvard \& Smithsonian, MS-78, 60 Garden St., Cambridge, MA 02138 USA}

\begin{abstract}

The Milky Way's Central Molecular Zone (CMZ) is the largest concentration of dense molecular gas in the Galaxy, the structure of which is shaped by the complex interplay between Galactic-scale dynamics and extreme physical conditions. Understanding the 3-D geometry of this gas is crucial as it determines the locations of star formation and subsequent feedback. We present a catalogue of clouds in the CMZ using Herschel data. Using archival data from the APEX and MOPRA CMZ surveys, we measure averaged kinematic properties of the clouds at 1mm and 3mm. We use archival ATCA data of the H$_{2}$CO (1$_{1,0}$ - 1$_{1,1}$) 4.8~GHz line to search for absorption towards the clouds, and 4.85~GHz GBT C-band data to measure the radio continuum emission. We measure the absorption against the continuum to provide new constraints for the line-of-sight positions of the clouds relative to the Galactic centre, and find a highly asymmetric distribution, with most clouds residing in front of the Galactic centre. The results are compared with different orbital models, and we introduce a revised toy model of a vertically-oscillating closed elliptical orbit. We find that most models describe the PPV structure of the gas reasonably well, but find significant inconsistencies in all cases regarding the near vs.\ far placement of individual clouds. Our results highlight that the CMZ is likely more complex than can be captured by these simple geometric models, along with the need for new data to provide further constraints on the true 3-D structure of the CMZ.

\end{abstract}


\section{Introduction}
\label{sec:intro}
The Milky Way's Central Molecular Zone (CMZ, inner few hundred parsecs) is home to the Galaxy's largest reservoir of dense molecular gas (total mass \textgreater \ 10$^{7}$~\msun, average density \textgreater \ 10$^{4}$~\cmq). With densities, gas temperatures, pressures, line-widths, cosmic ray ionisation rates, and magnetic field strengths that are several factors to orders of magnitude greater than in the Galactic disc, the CMZ provides a unique laboratory in which to study the environmental dependence of star formation on a range of physical scales that are inaccessible in extragalactic analogues \citep[see reviews by][]{Morris1996, Bryant2021, Henshaw2023}.

Mounting evidence shows that, under these conditions, the star formation rate in the CMZ is at least an order of magnitude lower than expected given the amount of dense gas \citep[e.g.][]{Longmore2013a, Barnes2017}. This is likely due to the inhibition of compact structure formation, within which star formation can occur, until much higher gas densities are reached \citep{Kauffmann2017b, Kauffmann2017c, Ginsburg2018b, Battersby2020}. Once star formation does occur, however, the process does not appear to fundamentally differ from that in Galactic disc regions \citep[e.g][]{Walker2021, Lu2020, Lu2021}.

\begin{figure*}
\includegraphics[width=\textwidth]{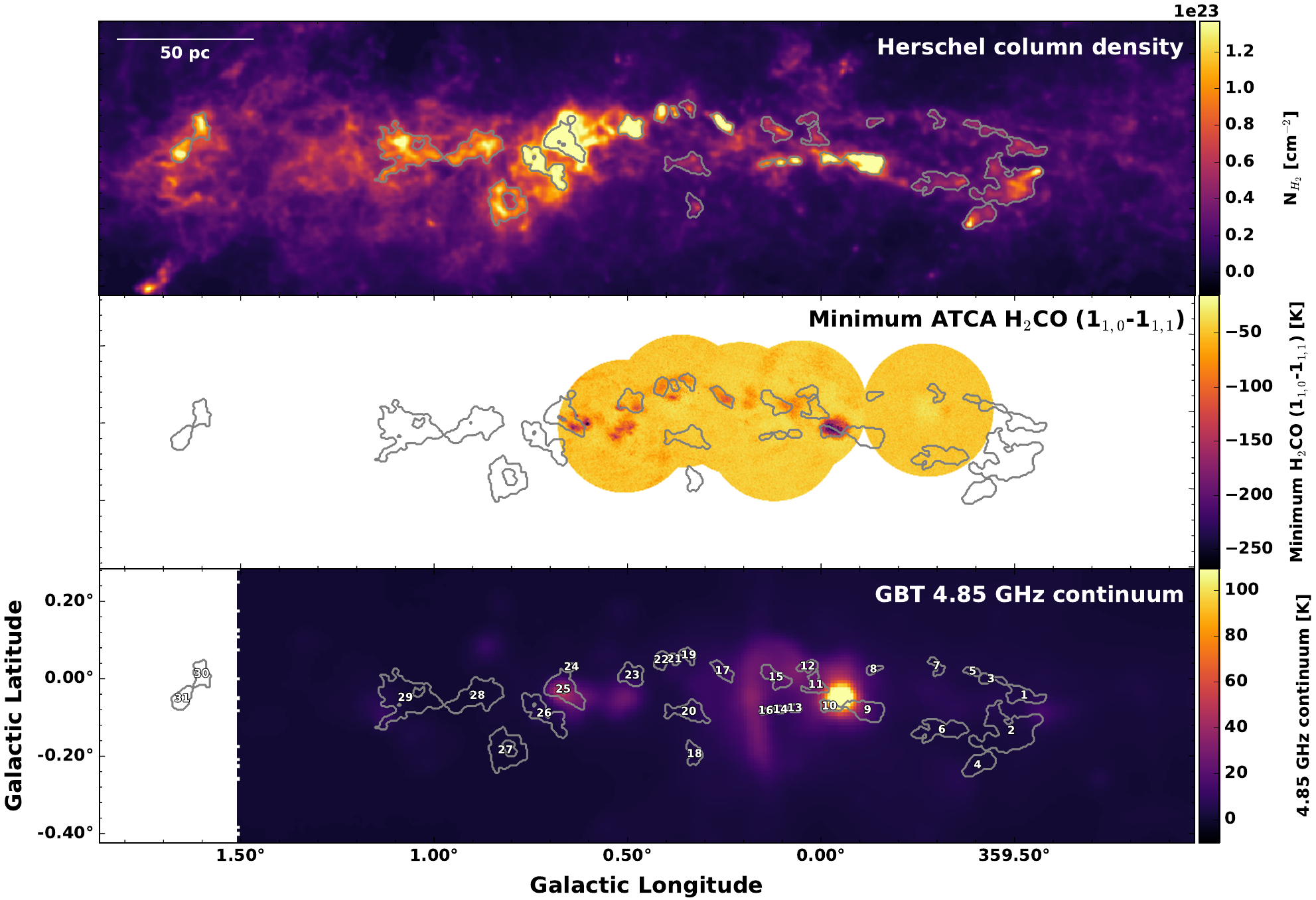}
\caption{Top: HiGAL column density map of the CMZ. Centre: ATCA H$_{2}$CO (1$_{1,0}-1_{1,1}$) minimum intensity map. Bottom: GBT C-Band continuum map. Contours show the top-level structures (leaves) identified using dendrograms. Numbers in the bottom panel correspond to the structure IDs in Table \ref{tab:leafgeneralproperties} and subsequent tables.}
\label{fig:colmap}
\end{figure*}

The majority of the dense gas and present-day star formation in the CMZ is confined to an apparent ring-like structure, which resembles a twisted, $\infty$-shaped ellipse in projection (Figure \ref{fig:colmap}). This structure has generally been viewed as some form of eccentric orbit with a Galactocentric radius \app 100~pc, and is therefore often referred to as the \q{100 pc stream} \citep{Molinari2011, Kruijssen2015}. This view is broadly accepted, and is supported by observations of more face-on extragalactic nuclei, which often also show ring- or spiral-like circumnuclear orbits on similar physical scales \citep[e.g.][]{Peeples2006, Comeron2010, Stuber2021, Sun2024}.

The exact 3-D geometry of our own CMZ, however, remains an open and debated topic. Understanding the 3-D structure is important, as it is intimately tied to significant Galactic-scale processes, such as the gas flows along the bar towards the Galactic centre, the distribution of dense gas and star formation, and the transport of material from the circumnuclear orbit towards Sgr A* \citep[e.g.][]{Ridley2017, Sormani2019a, Sormani2019b, Sormani2020b, Tress2020}.

The difficulty in observationally determining the 3-D structure of the CMZ is due to our perspective from within the Galactic plane. There is significant extinction along the line-of-sight, and we see the structure edge-on and have to rely on less direct measurements, such as position-position-velocity (PPV) information, to infer possible geometries. Using such methods, three main categories of potential geometries have emerged: (i) nuclear spiral arms \citep[e.g.][]{Sofue1995a, Ridley2017}, (ii) closed, vertically-oscillating elliptical orbits \citep[e.g.][]{Molinari2011, Sofue2022}, and (iii) an open, eccentric orbit that appears as a pretzel-shaped open gas stream in projection \citep{Kruijssen2015}. All of these models can be considered more generally as possible perturbations of x$_{\textrm{2}}$-like orbits \citep{Binney1991,Tress2020}.

\citet{Henshaw2016b, Henshaw2023} compared these three models and concluded that while the \citet{Molinari2011} elliptical orbit matches the projected $lb$ structure, it is a poor fit to the gas kinematics, confirming the conclusions of \citet{Kruijssen2015}. They conclude that both the spiral arms and open streams models provide a reasonable fit to the PPV data, but that there are significant disagreements in the locations of specific molecular clouds and star-forming regions, namely whether they are in front of or behind the Galactic centre (the location of Sgr A*).

A key piece of the puzzle in understanding the 3-D geometry of the CMZ is then to obtain stronger constraints on the line-of-sight positions of molecular clouds and, ultimately, the distances to them. Several studies have placed constraints on the positions and even distances of a handful of CMZ clouds \citep[e.g.][]{Yan2017, Nogueras-Lara2021}, but a comprehensive catalogue of the clouds and their relative positions and distances does not yet exist. In Battersby et al. (submitted), \textit{Papers I \& II} hereafter, data from the Herschel Infrared Galactic plane survey \citep[HiGAL,][]{Molinari2010} were used to produce dust column density and temperature maps of the inner 300~pc of the Galaxy. Using the dust column density map, dendrograms were used to characterise the full hierarchical structure and measure global properties of the dense gas in the CMZ. This represents the first step towards understanding the 3-D distribution of the CMZ gas.

In this paper, we focus on the peaks in this hierarchy -- the molecular clouds. Using the results from \textit{Papers I \& II} as a foundation, we produce a comprehensive catalogue of CMZ clouds and measure their physical and kinematic properties, and determine whether they are likely in the foreground or the background relative to the Galactic centre via radio absorption analysis. These results are then compared with various models of the 3-D geometry of the CMZ. An overview of the data used is given in Section \ref{sec:data}. The results are presented in Section \ref{sec:results}. A discussion of the results and the comparison with 3-D models of CMZ dense gas geometry is given in Section \ref{sec:discussion}. Conclusions are given in Section \ref{sec:conclusions}.

\section{Data}
\label{sec:data}

\subsection{Herschel dust column density}

The data that lay the foundation for this study come from the Herschel Infrared Galactic plane survey (HiGAL). Full details of the observations can be found in \citet{Molinari2010}. To summarise, the survey observed the Galactic plane from $\vert l \vert \leq  70^{\circ}$ and $\vert b \vert \leq \ 1^{\circ}$ at 70, 160, 250, 350, and 500~$\mu$m using the Spectral and Photometric Imaging Receiver (SPIRE) and the Photodetector Array Camera and Spectrometer (PACS). The corresponding beam sizes in each band are \app 6\asec, 12\asec , 18\asec , 25\asec , and 40\asec, respectively \citep{Traficante2011}.

In \textit{Paper I}, modified blackbody fits to the observed data were used to determine the spectral energy distribution (SED). This was then used to create a column density map of the CMZ, which is shown in Figure \ref{fig:colmap}. We refer the reader to \textit{Paper I} for a detailed description of the SED fitting and estimation of the dust column densities and temperatures. In this paper, we use the column density map to identify structures and produce a catalogue of CMZ clouds.

\subsection{MOPRA 3~mm CMZ line survey}
\label{subsec:mopra}

To measure kinematic properties of the clouds, we utilise data from the MOPRA 3~mm CMZ survey. We use data from the 83 -- 95~GHz portion of the survey \citep{Jones2012}. The angular resolution of the survey is 39\asec, and the spectral resolution is 2~\kms. A total of 15 molecular tracers were observed, spanning a range in excitation conditions and critical densities. For our purposes, we choose to focus on 3 specific tracers, namely HNCO (4$_{0,4}$ - 3$_{0,3}$), HCN (1-0), and HC$_{3}$N (10-9). We select these lines as they are typically used as \q{dense gas} tracers. These lines are therefore well-suited to our goals, as we are interested primarily in the densest parts of the CMZ gas -- the molecular clouds. In particular, the HNCO (4$_{0,4}$ - 3$_{0,3}$) emission from this survey has been shown to reliably trace the underlying kinematics of the dense gas in the CMZ \citep{Henshaw2016b}.

\subsection{APEX 1~mm CMZ line survey}
\label{subsec:apex}

In addition to the 3~mm molecular line data, we also use the 1~mm data from the APEX line survey of the CMZ \citep{Ginsburg2016}. The angular resolution of the survey is 30\asec, and the spectral resolution is 1~\kms. The primary target lines for this survey were the para-H$_{2}$CO transitions at \app 218~GHz, along with the $^{13}$CO and C$^{18}$O (2-1) transitions. There were many other lines detected in the spectral windows, however we choose to focus only on the aforementioned primary target lines. In particular, we use the H$_{2}$CO (3$_{0,3}$ - 2$_{0,2}$) line as our primary tracer at 1~mm, as it most reliably traces the dense gas that we are interested in. The higher excitation para-H$_{2}$CO lines have lower signal-to-noise, and while the CO isotopologues are well detected, they are abundant throughout the CMZ and along the line-of-sight, complicating their spectra (e.g. Figure \ref{fig:multispec}).

\subsection{ATCA \& GBT data}
We use archival data from the ATCA CMZ survey (PI: Adam Ginsburg, project C3045\footnote{The data from this survey can be found at \url{https://doi.org/10.7910/DVN/R08UCE}.}). Specifically, we use the 1$_{1,0}-1_{1,1}$ transition of H$_{2}$CO at 4.82966~GHz to search for absorption signatures towards CMZ clouds. The velocity resolution of the survey is 1.9~\kms, and the synthesised beam is 4.84\asec $\times$ 1.49\asec. See \citet{Ginsburg2015} for further survey details. The footprint of this survey covers most of the CMZ, but unfortunately it misses a few significant regions, including Sgr C, and portions of Sgr B2 and the 20~\kms cloud (Figure \ref{fig:colmap}).

We also use data from the GBT CMZ pilot survey\footnote{The data for the GBT CMZ pilot survey can be found at \url{https://dataverse.harvard.edu/dataset.xhtml?persistentId=doi:10.7910/DVN/I2U8GK}, and the data reduction information can be found at \url{https://github.com/keflavich/GBTLimaBrickH2CO}} to search for the absorption of the same H$_{2}$CO 4.82966~GHz line. However, this pilot survey only covers a portion of the CMZ, namely the dust ridge, which spans from G0.253+0.016 (the Brick) to Sagittarius B2. We therefore only use these data to verify the results from the ATCA data, and do not use them for analysis.

Finally, we use the GBT CMZ C-band (4.85~GHz) continuum maps from \citet{Law2008} to measure the radio continuum emission towards CMZ clouds. While the angular resolution of these data (2.5$^{\prime}$) is considerably coarser than the other data sets used here, it is the only publicly available 4.8~GHz continuum map that covers the entire CMZ. The coverage is shown in Figure \ref{fig:colmap}. Where direct comparisons between the ATCA and GBT data are made (e.g. Section \ref{sec:fractional_abs}), the ATCA data are reprojected onto the same pixel grid as the GBT.

\section{Results}
\label{sec:results}

\subsection{Cloud catalogue}
\label{sec:catalogue}

In \textit{Paper II}, we used the Herschel column density map shown in Figure \ref{fig:colmap} to produce a catalogue of the structures in the map. This was done using dendrograms, specifically with the \texttt{astrodendro} Python package\footnote{\url{https://dendrograms.readthedocs.io/en/stable/}}. To briefly summarise, dendrograms characterise the hierarchical structure in data, which is represented as a \q{tree}, where substructures are classified as \q{branches}, and local maxima at the highest level of the branch structures are called \q{leaves}.

We previously reported the full hierarchy of the dendrogram in \textit{Paper II}, which used a single set of parameters to identify the structure of all of the dense gas in the CMZ. In this paper, we are explicitly interested in the top-level structures (the leaves), which in this context we assume to represent individual molecular clouds. Initially, we compute the dendrogram using the same parameters as in \textit{Paper II}, namely a minimum threshold of 2$\times$10$^{22}$~\cms, an increment between structures of 5$\times$10$^{22}$~\cms, and a minimum number of pixels of 20.  Due to the hierarchical nature of the interstellar medium (ISM), the resulting dendrogram is influenced significantly by the choice of these parameters. Our choice of parameters is ultimately guided by prior knowledge of certain \q{clouds} that are already known in the CMZ (e.g. \q{the Brick}, the dust ridge clouds, the 20 \kms and 50 \kms clouds, etc.), such that they are identified as leaves in the dendrogram.

Given the significant dynamic range in column densities in the CMZ, we find that a single dendrogram with a single set of parameters only identifies the most prominent clouds as leaves, namely those in the upper left and lower right quadrant of the CMZ \q{ring} (see Figure \ref{fig:colmap}). This is not surprising, but for our purposes of identifying clouds throughout the ring and comparing with 3-D models of the gas geometry, it is crucial that we capture more of the lower density clouds, which still represent maxima in contrast to their local surroundings.

To overcome this limitation, we treat the upper right and lower left quadrants of the $\infty$-shaped structure individually, and compute dendrograms for these regions separately. This is done by masking the map outside of these quadrants, which are identified visually. We also lower the increment between structures from 5$\times$10$^{22}$~\cms\ to 2$\times$10$^{22}$~\cms, as the density contrast in these regions is lower. The dendrogram threshold is also increased to 3$\times$10$^{22}$~\cms, and the minimum number of pixels is unchanged. Similarly, for the \q{three little pigs} \citep{Battersby2020}, we decrease the increment to 1$\times$10$^{22}$~\cms due to the low density contrast between the clouds. The threshold and number of pixels are unchanged.

We also explicitly change one of the leaves. In the original catalogue, G0.253+0.016 (aka \q{the Brick}) was identified as a much more extended structure than is typically presented in the literature. Visual inspection of the molecular line emission discussed in the following subsections shows that this extended region of the cloud is kinematically distinct. We therefore explicitly force a higher threshold of 7$\times$10$^{22}$~\cms \ for this region in order to isolate the known extent of the cloud.

The leaf catalogues generated from these multiple dendrograms are then concatenated into a single catalogue representing all of the clouds identified. We then prune the catalogue to remove specific leaves that are identified at much lower and higher latitudes (-0.25 \textless\ $b$ \textless\ 0.1$^\circ$). The reasons for removing these leaves are (i) it is not clear that they actually reside in the CMZ, and (ii) they show little-to-no significant line emission in any of the tracers used in this paper. The concatenated and pruned cloud catalogue is then re-indexed according to Galactic longitude, increasing from negative to positive longitudes. As this procedure for structure identification differs from that used in \textit{Paper II}, the leaves in the two catalogues are not identical. We refer the reader to Table 1 in \textit{Paper II}, which provides cross-matching between the two catalogues, where applicable.

The resulting cloud catalogue is shown via contours in Figure \ref{fig:colmap}, with numbered structure indices overlaid. A total of 31 clouds are identified. Table \ref{tab:leafgeneralproperties} presents a complete overview of the catalogue, along with various physical and kinematic properties, the measurements of which are described in the following subsections.

\subsection{Cloud physical properties}

As the input map is already in units of column density, it is straightforward to extract physical quantities of the clouds. Median and maximum column densities are computed directly from the dendrogram catalogue. The mask for each leaf is then mapped onto the corresponding Herschel dust temperature maps (see \textit{Paper I}) to extract median and maximum dust temperatures within the clouds. The column density of each cloud is then used to determine the mass via

\begin{equation}
\label{eq:mass}
M = \mu_{\mathrm{H_{2}}} m_{\mathrm{H}} \int N_{\mathrm{H_{2}}} \mathrm{d}A
\end{equation}

\noindent where $M$ is the mass, $\mu_{\mathrm{H_{2}}}$ is the mean molecular weight, assumed to be 2.8, $m_{\mathrm{H}}$ is the mass of atomic hydrogen, $N_{\mathrm{H_{2}}}$ is the H$_{2}$ column density, and A is the area. A distance of 8.1~kpc is assumed for all clouds to estimate their areas, radii, and masses, which is the distance to the Galactic centre \citep{GravityCollaboration2019}. The actual distances to individual clouds will vary from this value, but given that we only expect variations of $\pm$ 100~pc (1-2\%), this serves as a reasonable estimate.

We note that the measured column densities, masses, and dust temperatures are dependent on a number of assumptions. In particular, we assume a fixed dust opacity index, $\beta$ of 1.75 in \textit{Paper I}. However, \citet{Tang2021b, Tang2021a} estimated $\beta$ throughout the CMZ by combining AzTEC LMT data with Bolocam, Plack, and Herschel data. Their results suggest that $\beta$ ranges from 2.0 -- 2.4, showing a positive correlation with column density. This means that their reported column densities and dust temperatures are slightly different than what we report here and in \textit{Paper I}. Despite this, we find that the column densities and dust temperatures towards the clouds in our catalogue are broadly consistent with those presented in \citet{Tang2021b}, with typical differences \app 12\%, and as much as 25\%. The median temperature and column density errors reported in \textit{Paper I} are 11\% and 25\%, respectively. The differences with the results from \citet{Tang2021b} are therefore well within the systematic uncertainties.

Overall, we find that the clouds range in mass from \app 1$\times$10$^{4}$ -- 1$\times$10$^{6}$~\msun, with a median of 6$\times$10$^{4}$~\msun. Radii range from 1.4 -- 8.7~pc, with a median of 3.4~pc, and are calculated as the effective radius of a circle with an area equivalent to that of the source. Averaged dust temperatures are 15 -- 29~K, with a median of 22~K. Averaged column densities are 2.8$\times$10$^{22}$ -- 3.6$\times$10$^{23}$~\cms, with a median of 8.7$\times$10$^{22}$~\cms (see Table \ref{tab:leafgeneralproperties}).

\begin{deluxetable*}{cccccccccccccccc}
\tablecaption{Properties of the cloud catalogue}
\rotate
\tablewidth{0pt}
\tablehead{
\colhead{\#} & \colhead{Area} & \colhead{$l$} & \colhead{$b$} & 
\colhead{Median} & \colhead{Peak} & \colhead{Mass} & \colhead{Radius} & 
\colhead{Median} & \colhead{Peak} & \colhead{$v_{\textrm{HNCO}}$} & 
\colhead{$\sigma_{\textrm{HNCO}}$} & \colhead{$v_{\textrm{H$_{2}$CO}}$} & 
\colhead{$\sigma_{\textrm{H$_{2}$CO}}$} & \colhead{Colloquial} \\
\colhead{} & \colhead{} & \colhead{} & \colhead{} & 
\colhead{N$_{\textrm{H$_{2}$}}$} & \colhead{N$_{\textrm{H$_{2}$}}$} & 
\colhead{} & \colhead{} & \colhead{T$_{d}$} & \colhead{T$_{d}$} & 
\colhead{} & \colhead{} & \colhead{} & \colhead{} & \colhead{name} \\
\hline
\colhead{} & \colhead{pc$^{2}$} & \colhead{$^{\circ}$} & \colhead{$^{\circ}$} & 
\colhead{cm$^{-2}$} & \colhead{cm$^{-2}$} & \colhead{M$_{\odot}$} & \colhead{pc} & 
\colhead{K} & \colhead{K} & \colhead{km s$^{-1}$} & \colhead{km s$^{-1}$} & 
\colhead{km s$^{-1}$} & \colhead{km s$^{-1}$} & \colhead{}
}
\startdata
1 & 57 & -0.525 & -0.044 & 4.7E+22 & 6.7E+22 & 6.0E+04 & 4.3 & 23 & 26 & -102 & 6 & -102 & 5 & - \\
2 & 236 & -0.492 & -0.135 & 4.9E+22 & 1.9E+23 & 2.9E+05 & 8.7 & 24 & 28 & -56 & 9 & -55 & 8 & Sgr C \\
3 & 22 & -0.439 & -0.001 & 4.1E+22 & 5.4E+22 & 2.1E+04 & 2.7 & 24 & 26 & -90 & 8 & -91 & 5 & - \\
4 & 66 & -0.405 & -0.223 & 4.8E+22 & 1.6E+23 & 8.1E+04 & 4.6 & 23 & 25 & -27, 20 & 13, 8 & 19 & 2 & - \\
5 & 13 & -0.392 & 0.018 & 4.2E+22 & 5.2E+22 & 1.2E+04 & 2.0 & 23 & 25 & -78 & 5 & -77 & 6 & - \\
6 & 97 & -0.312 & -0.132 & 3.7E+22 & 8.2E+22 & 8.7E+04 & 5.6 & 24 & 26 & -29, -21 & 5, 12 & -28 & 5 & - \\
7 & 18 & -0.299 & 0.032 & 3.4E+22 & 4.5E+22 & 1.4E+04 & 2.4 & 25 & 26 & -73, -37 & 11, 5 & -35 & 4 & - \\
8 & 12 & -0.135 & 0.023 & 3.6E+22 & 5.6E+22 & 1.1E+04 & 2.0 & 26 & 28 & -54, 15, 62 & 9, 15, 6 & -48, 64 & 5, 5 & - \\
9 & 78 & -0.120 & -0.081 & 1.6E+23 & 4.2E+23 & 3.1E+05 & 5.0 & 21 & 26 & 15 & 12 & 15 & 12 & 20 km/s \\
10 & 22 & -0.021 & -0.071 & 1.2E+23 & 1.8E+23 & 6.2E+04 & 2.7 & 24 & 25 & 48 & 11 & 46 & 13 & 50 km/s \\
11 & 37 & 0.014 & -0.016 & 4.4E+22 & 7.1E+22 & 3.8E+04 & 3.4 & 29 & 32 & -11, 45, 14 & 9, 9, 8 & -11, 51 & 9, 9 & - \\
12 & 22 & 0.035 & 0.032 & 4.2E+22 & 6.6E+22 & 2.2E+04 & 2.7 & 29 & 32 & 86 & 7 & 86 & 8 & - \\
13 & 12 & 0.068 & -0.076 & 1.1E+23 & 1.6E+23 & 3.0E+04 & 1.9 & 21 & 22 & 50 & 9 & 51 & 10 & Stone \\
14 & 9 & 0.105 & -0.080 & 1.0E+23 & 1.4E+23 & 2.2E+04 & 1.7 & 22 & 23 & 53 & 9 & 54 & 9 & Sticks \\
15 & 49 & 0.116 & 0.003 & 5.3E+22 & 1.1E+23 & 6.2E+04 & 4.0 & 25 & 28 & 52 & 11 & 52 & 11 & - \\
16 & 8 & 0.143 & -0.083 & 8.9E+22 & 1.0E+23 & 1.5E+04 & 1.6 & 22 & 23 & -15, 57 & 3, 8 & -15, 58 & 3, 10 & Straw \\
17 & 31 & 0.255 & 0.020 & 1.4E+23 & 2.7E+23 & 1.0E+05 & 3.1 & 21 & 24 & 18, 37, 70 & 16, 7, 14 & 14, 36 & 8, 15 & Brick \\
18 & 33 & 0.327 & -0.195 & 2.8E+22 & 7.1E+22 & 2.4E+04 & 3.2 & 22 & 27 & 16 & 10 & - & - & - \\
19 & 20 & 0.342 & 0.060 & 4.8E+22 & 9.9E+22 & 2.5E+04 & 2.6 & 23 & 26 & -2 & 13 & 1 & 10 & B \\
20 & 70 & 0.342 & -0.085 & 3.8E+22 & 6.9E+22 & 6.3E+04 & 4.7 & 25 & 28 & 90 & 18 & 94 & 11 & Sailfish \\
21 & 9 & 0.379 & 0.050 & 8.7E+22 & 1.3E+23 & 1.9E+04 & 1.7 & 21 & 24 & 8, 39 & 14, 4 & 8, 39 & 9, 3 & C \\
22 & 26 & 0.413 & 0.048 & 9.7E+22 & 2.0E+23 & 6.2E+04 & 2.9 & 21 & 23 & 19 & 11 & 18 & 10 & D \\
23 & 54 & 0.488 & 0.008 & 1.4E+23 & 3.9E+23 & 2.0E+05 & 4.1 & 20 & 22 & 28 & 11 & 31 & 10 & E/F \\
24 & 6 & 0.645 & 0.030 & 2.0E+23 & 2.3E+23 & 2.8E+04 & 1.4 & 19 & 20 & 53 & 12 & 50 & 13 & - \\
25 & 104 & 0.666 & -0.028 & 3.6E+23 & 2.1E+24 & 1.1E+06 & 5.8 & 20 & 25 & 62 & 12 & 52, 70 & 6, 9 & Sgr B2 \\
26 & 101 & 0.716 & -0.090 & 1.4E+23 & 2.3E+23 & 3.2E+05 & 5.7 & 20 & 22 & 28, 58 & 15, 11 & 25, 59 & 14, 7 & G0.714 \\
27 & 128 & 0.816 & -0.185 & 9.0E+22 & 1.4E+23 & 2.7E+05 & 6.4 & 19 & 20 & 39 & 15 & 42 & 11 & - \\
28 & 147 & 0.888 & -0.044 & 9.1E+22 & 1.4E+23 & 3.0E+05 & 6.8 & 19 & 20 & 14, 26, 84 & 11, 30, 8 & 9, 84 & 11, 7 & - \\
29 & 229 & 1.075 & -0.049 & 8.5E+22 & 1.7E+23 & 4.6E+05 & 8.5 & 18 & 23 & 74, 85 & 16, 6 & 80 & 10 & - \\
30 & 45 & 1.601 & 0.012 & 9.7E+22 & 1.7E+23 & 1.0E+05 & 3.8 & 16 & 17 & 48, 58 & 5, 11 & 54 & 6 & G1.602 \\
31 & 43 & 1.652 & -0.052 & 1.0E+23 & 1.8E+23 & 1.0E+05 & 3.7 & 15 & 16 & 50 & 6 & 52 & 5 & G1.651 \\ \hline 
\enddata
\tablecomments{Shown for each source is the exact leaf area in parsec$^{2}$, central coordinates in degrees ($l, b$), median and peak column density (N$_{\textrm{H$_{2}$}}$), mass, radius, median and peak dust temperature (T$_{d}$), central velocity ($v$) and velocity dispersion from Gaussian fitting ($\sigma$) for both HNCO and H$_{2}$CO, and colloquial name in the literature. Where multiple velocity components are observed, the central velocities and dispersions of each are given.}
\label{tab:leafgeneralproperties}
\end{deluxetable*}

\subsection{Cloud kinematic properties}
\label{sec:kinematic_properties}

To measure averaged kinematic properties of the clouds in our catalogue, we explored a total of eight molecular lines -- three at 3~mm and five at 1~mm (Sections \ref{subsec:mopra} \& \ref{subsec:apex}). For every cloud in the catalogue, we produced an individual mask corresponding to the area of the dendrogram leaf. Using these masks, we extract spatially-averaged spectra, integrated intensity maps, velocity maps (moment 1), and velocity dispersion maps (moment 2) for each of the eight lines for all clouds. We note that these are naive moment maps that do not account for multiple velocity components. These are accounted for later when performing multi-component Gaussian fitting (see also Appendix \ref{ap:gauss}). 

\begin{figure*}
\includegraphics[width=\textwidth]{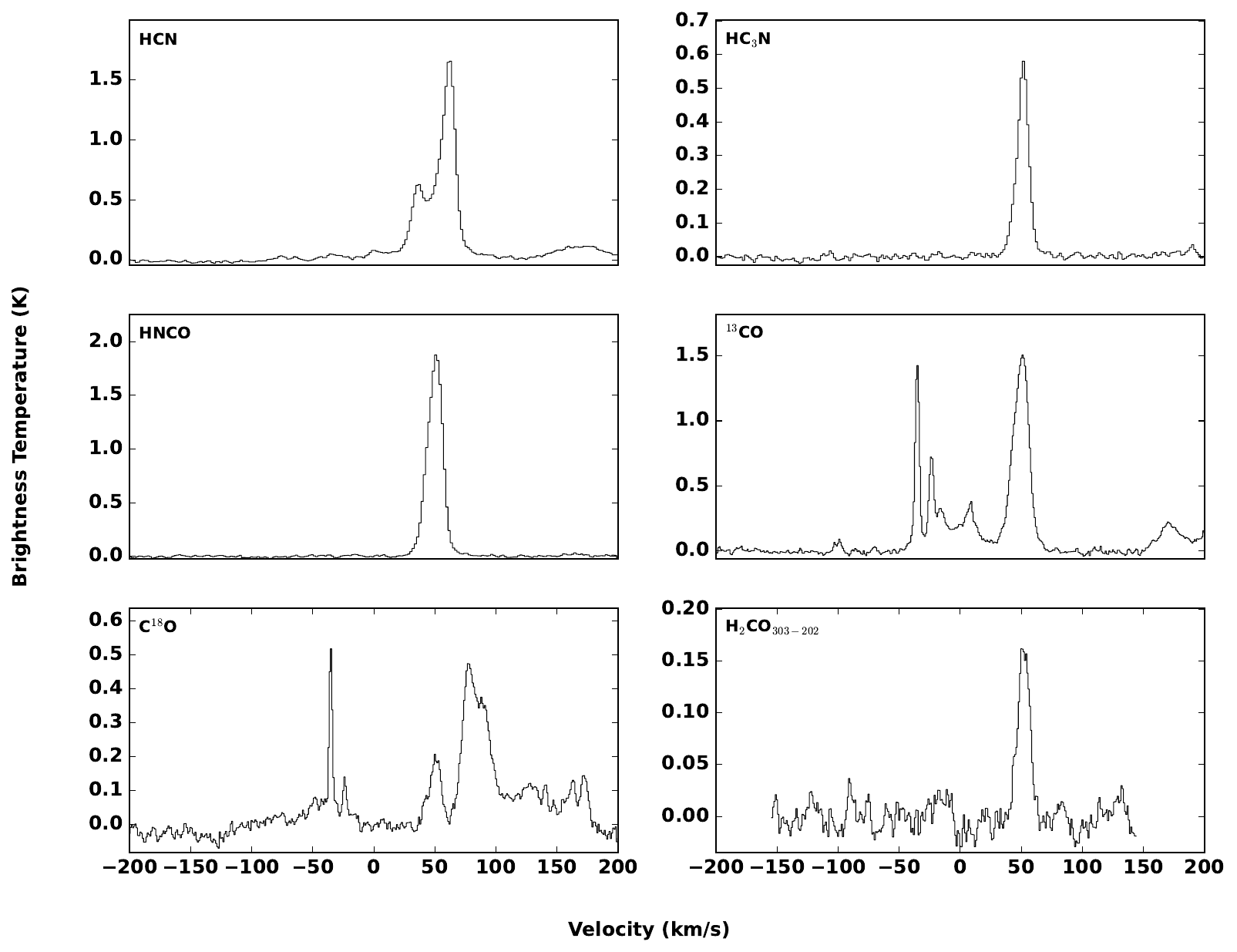}
\caption{Averaged spectra towards source 31 for all six of the molecular line tracers that we considered. Note that the y-axes are scaled independently for clarity.}
\label{fig:multispec}
\end{figure*}

An example of the averaged spectra towards source 31 (G1.651) is shown in Figure \ref{fig:multispec}. The two higher energy transitions of para-H$_{2}$CO are not shown, as the signal-to-noise is low. This example shows a wide variety of spectral complexity, notably for HCN, $^{13}$CO, and C$^{18}$O. This is commonly observed towards all clouds, which is not unexpected, as these molecules are very abundant both in the CMZ and along the line-of-sight. 

\begin{figure*}
\includegraphics[width=\textwidth]{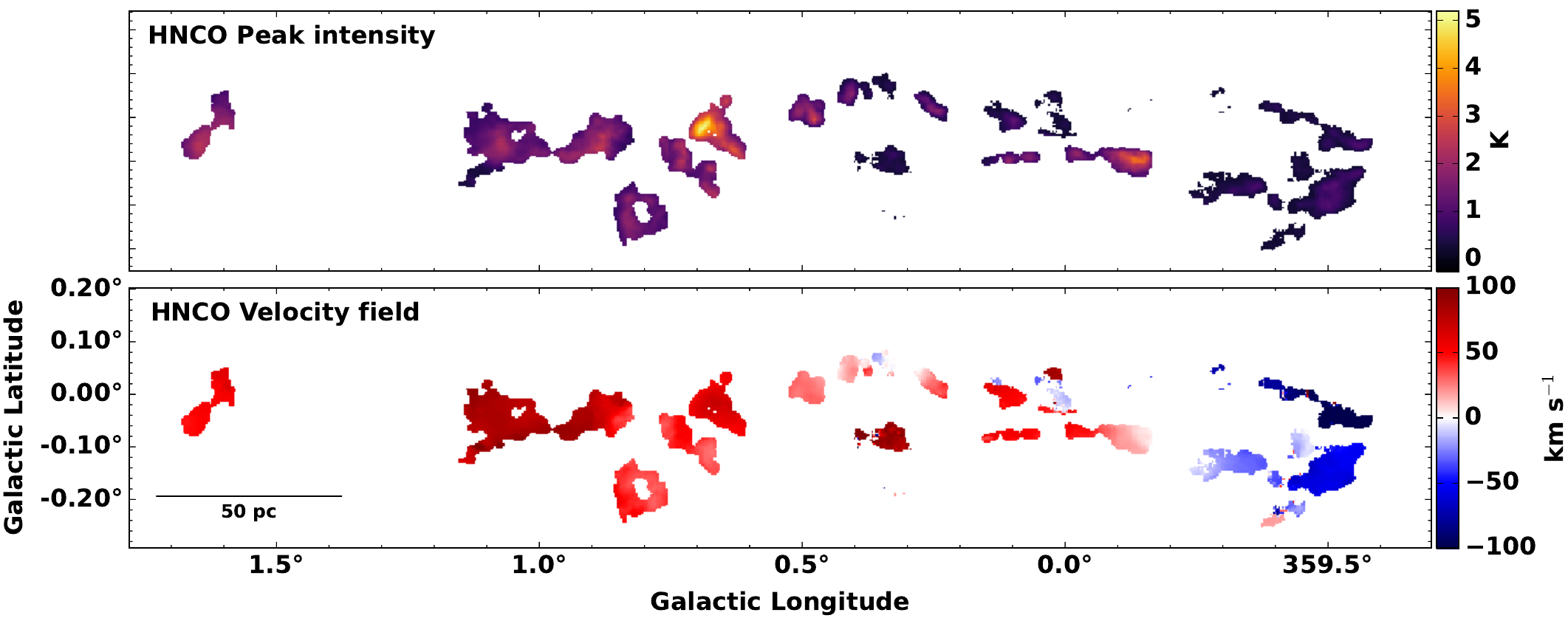}
 \caption{Overview of the HNCO emission within the dendrogram leaves determined from the HiGAL dust continuum emission. Shown are maps of the peak intensity (top), and the velocity field (bottom). Note that this is a naive first moment map, and does not account for multiple velocity components (see Section \ref{sec:kinematic_properties}).}
\label{fig:mom_hnco}
\end{figure*}

The remaining molecular lines are those from HC$_{3}$N, HNCO, and H$_{2}$CO. These are all tracers of denser gas, and so in principle they should more accurately trace the underlying cloud kinematics compared to the other lines. From the group of lines at 3mm, we choose the HNCO (4$_{0,4}$ - 3$_{0,3}$) as our primary kinematic tracer, as it is typically brighter than HC$_{3}$N. HNCO (4$_{0,4}$ - 3$_{0,3}$) has also been shown to reliably trace the dense gas throughout the CMZ \citep{Henshaw2016b, Henshaw_2019}. From the lines at 1mm, we choose H$_{2}$CO (3$_{0,3}$ - 2$_{0,2}$). Though it is comparatively faint, it is the brightest of the three para-H$_{2}$CO transitions, and the spectra from the CO isotopologues are too complex. H$_{2}$CO emission at 218~GHz also traces the dense, warm gas throughout the CMZ \citep{Ginsburg2016}. Figure \ref{fig:mom_hnco} shows an overview of the HNCO emission throughout our CMZ cloud catalogue.

Having selected our two main kinematic tracers, we fit the averaged HNCO and H$_{2}$CO spectra for all clouds in the catalogue. This is done using multiple component Gaussian fitting using \texttt{pyspeckit}, which uses the Levenberg-Marquardt algorithm. For each cloud, we provide an initial guess for the number of components (which is determined by-eye), and the amplitude, central velocity, and FWHM of each component. \texttt{pyspeckit} generally does a good job of fitting the spectra with rough initial guesses, but in the few cases where fitting is poor, we provide revised input parameters and redo the fitting. 

The best-fitting central velocities and velocity dispersions for the averaged HNCO and H$_{2}$CO are presented in Table \ref{tab:leafgeneralproperties}. We note that while there is generally good agreement between the results from fitting HNCO and H$_{2}$CO, there are discrepancies. Most notably, there are several clouds for which the HNCO spectra show additional velocity components that are not obviously seen in H$_{2}$CO. This may be due to the fact that the HNCO is generally a factor of a few to an order of magnitude brighter, and so fainter components are better detected, but it could also be that the two lines are not tracing exactly the same cloud material.

Based on the averaged HNCO emission, the central velocities of the cloud components range from \app -100 to +90~\kms, and velocity dispersions range from 3 -- 30~\kms, with a median of 10~\kms. Of the 31 clouds in the catalogue, we find that 19 are well described by a single Gaussian component, 8 by two components, and 4 by three components. Note that this is a simplistic approach to capture the average kinematic properties of the clouds, namely their central velocities and velocity dispersions. Detailed spectral decomposition on a pixel-by-pixel basis shows that the velocity structure in CMZ clouds is considerably more complex \citep[e.g.][]{Henshaw2016b, Henshaw_2019}. However, this is beyond the scope of what is needed for the analyses presented in this work.

\subsection{Formaldehyde 4.8~GHz absorption}
\label{sec:h2co_absorption}

Formaldehyde (H$_{2}$CO) is effective at absorbing at centimetre wavelengths, and can even be seen in absorption against the Cosmic Microwave Background (CMB). By searching for deep absorption that is correlated both spatially and kinematically, we aim to use the strength of H$_{2}$CO absorption towards the clouds in our catalogue to constrain whether they are likely to be in front of or behind the Galactic centre. The underlying assumption here is that the Galactic centre is filled with diffuse ionised gas, and that a cloud on the near side of the CMZ will show absorption deeper than 2.73~K, whereas a cloud on the far side will not.  

Our primary tracer for absorption is the H$_{2}$CO (1$_{1,0}-1_{1,1}$) 4.82966~GHz line. There are no publicly available high-resolution maps of the line that cover all of the clouds in our catalogue. The best available data come from the ATCA CMZ survey, which observed the majority of the inner CMZ at \app 3\asec resolution \citep{Ginsburg2015}. The footprint ranges from 0.795 -- 359.436 in longitude, and -0.674 -- +0.624 in latitude, effectively covering catalogue structures 3 -- 25, although structures 4 and 18 lie just outside of the latitude coverage of the survey. Of the 31 clouds in our catalogue, 21 have full or partial coverage with the ATCA data (see Figure \ref{fig:colmap}).

For each of the 21 clouds for which we have corresponding H$_{2}$CO (1$_{1,0}-1_{1,1}$) data, we create spatially-cropped sub-cubes within the corresponding dendrogram masks, then extract spatially-averaged spectra. We find that all of the regions show some level of absorption, ranging from -3 to -100~K. 

To determine whether the absorption is physically associated with a cloud, we investigate several diagnostics. We compare the averaged H$_{2}$CO (1$_{1,0}-1_{1,1}$) absorption with the HNCO emission at 3mm and H$_{2}$CO emission at 1~mm to check whether they occur at the same velocities. We then compare the morphology of the integrated H$_{2}$CO (1$_{1,0}-1_{1,1}$) intensity with that of the H$_{2}$CO and HNCO integrated emission. We also take position-velocity slices along the major axes of the clouds to search for correlated absorption. 

\begin{figure*}
\includegraphics[width=\textwidth]{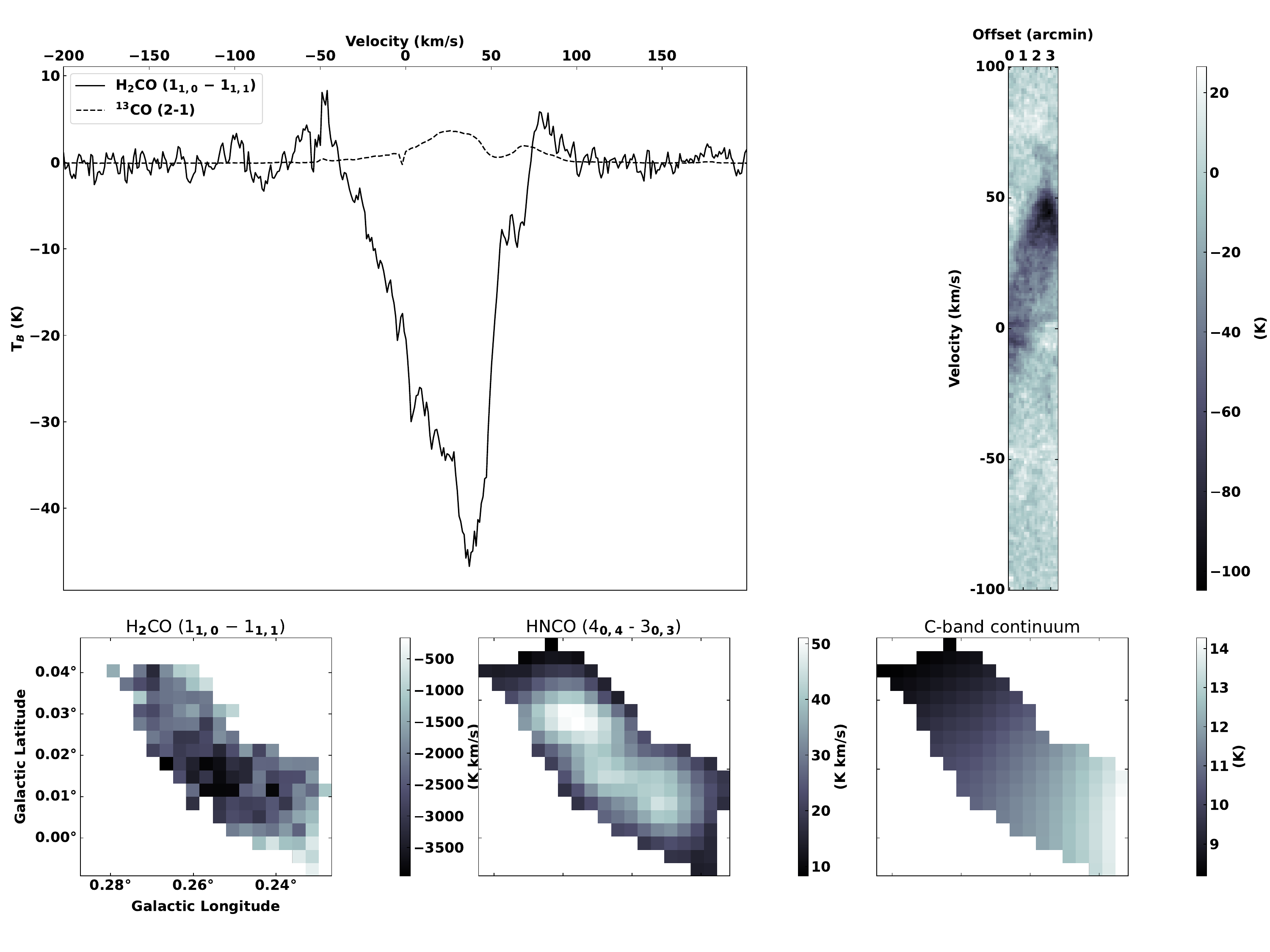}
\caption{Overview of the emission and absorption towards source 17 (\q{the Brick}). \textit{Top left}: Cloud-averaged spectra of $^{13}$CO (dashed black line) and H$_{2}$CO (1$_{1,0}-1_{1,1}$) (solid black line). \textit{Top right}: PV slice of H$_{2}$CO (1$_{1,0}-1_{1,1}$) taken along the major axis of the cloud. \textit{Bottom left}: Integrated H$_{2}$CO (1$_{1,0}-1_{1,1}$) intensity. \textit{Bottom centre}: Integrated HNCO (4$_{0,4}$ - 3$_{0,3}$) intensity. \textit{Bottom right}: C-band (4.85~GHz) continuum.}
\label{fig:absorption_brick}
\end{figure*}

The emission and absorption profiles of the clouds are often complex, but in general, where we see significant absorption, it is well correlated with the velocity of the emission at mm wavelengths, and correlated in PV-space. An example of this is shown in Figure \ref{fig:absorption_brick}. There are also cases where things are more ambiguous, in particular near to Sgr A* and Sgr B2 there is apparent strong H$_{2}$CO (1$_{1,0}-1_{1,1}$) emission. In general, such emission is not expected to be observed, and is likely an artifact in the data caused by positive sidelobes that can accompany strong absorption features (i.e. the inverse of \q{negative bowls} that are commonly observed near bright emission in interferometric data). There are also cases of genuine emission, most notably in the dust ridge cloud C (G0.038+0.04), where there is a bright H$_{2}$CO maser \citep{Ginsburg2015}.

The maximum absorption depth (minimum intensity) for each cloud is measured from the spatially-averaged spectrum, and is restricted such that the depth is measured within the dispersion about the centroid velocity of the molecular line emission. Where there are multiple velocity components seen in emission, this process is repeated within the dispersion about each velocity component, and the source IDs are split (e.g. 1a, 1b). The measured depths are given in Table \ref{tab:absorption}, along with other measurements that are described in the following subsections.

\subsection{Absorption against the centimetre continuum}
\label{sec:fractional_abs}

Centimetre continuum emission is widespread throughout the CMZ \citep[e.g.][]{Law2008, Heywood2022}. In addition to simply measuring the absorption depth, we also use the relative strength of H$_{2}$CO absorption against the cm-continuum background towards the clouds to provide additional constraints on their line-of-sight position relative to the Galactic centre.

To achieve this we use the C-band (4.85~GHz) continuum data from the GBT CMZ survey \citep{Law2008}. This survey covers all of the clouds in our catalogue, though because we are comparing directly to the line absorption, we only consider the 21 sources for which we have the ATCA data. For each cloud we measure the median continuum intensity to obtain a representative value of the continuum level towards the cloud.

In cases where a cloud has multiple velocity components, we explored several methods to account for this when measuring the median continuum emission. Initially, we opted to use the morphology of the integrated HNCO emission intensity maps to create spatial sub-masks within which to measure the continuum level. In practice this was difficult to implement due to the coarse angular resolution of the data and significant overlap between the multiple components. Ultimately we decided to implement two different methods based on the peak HNCO intensity. For each velocity component, we create a peak intensity map of the HNCO emission within a velocity range defined by the velocity dispersion of the component about the centroid velocity. Using these maps, the following techniques are then used.

\begin{itemize}
    \setlength{\itemindent}{0.5em}
    \item \textit{Peak intensity masking}: for a given pixel within the dendrogram mask for a given source, the values of the peak intensities of each component are compared. The pixel in the sub-mask corresponding to the velocity component with the greatest peak intensity is assigned a value of 1, while those in the remaining components are assigned NaN.
    \\
    \item \textit{Peak intensity weighting}: at each pixel within the sub-masks for each velocity component, the pixel is assigned a value corresponding to the fractional value of the peak intensity of that component relative to the sum of the peak intensities of all components. This effectively assigns a weight to every pixel for each component, such that the value for one component is between 0-1, and the sum of all sub-masks at a given pixel is equal to 1. 
\end{itemize}

\begin{figure}
\centering
\includegraphics[width=\columnwidth]{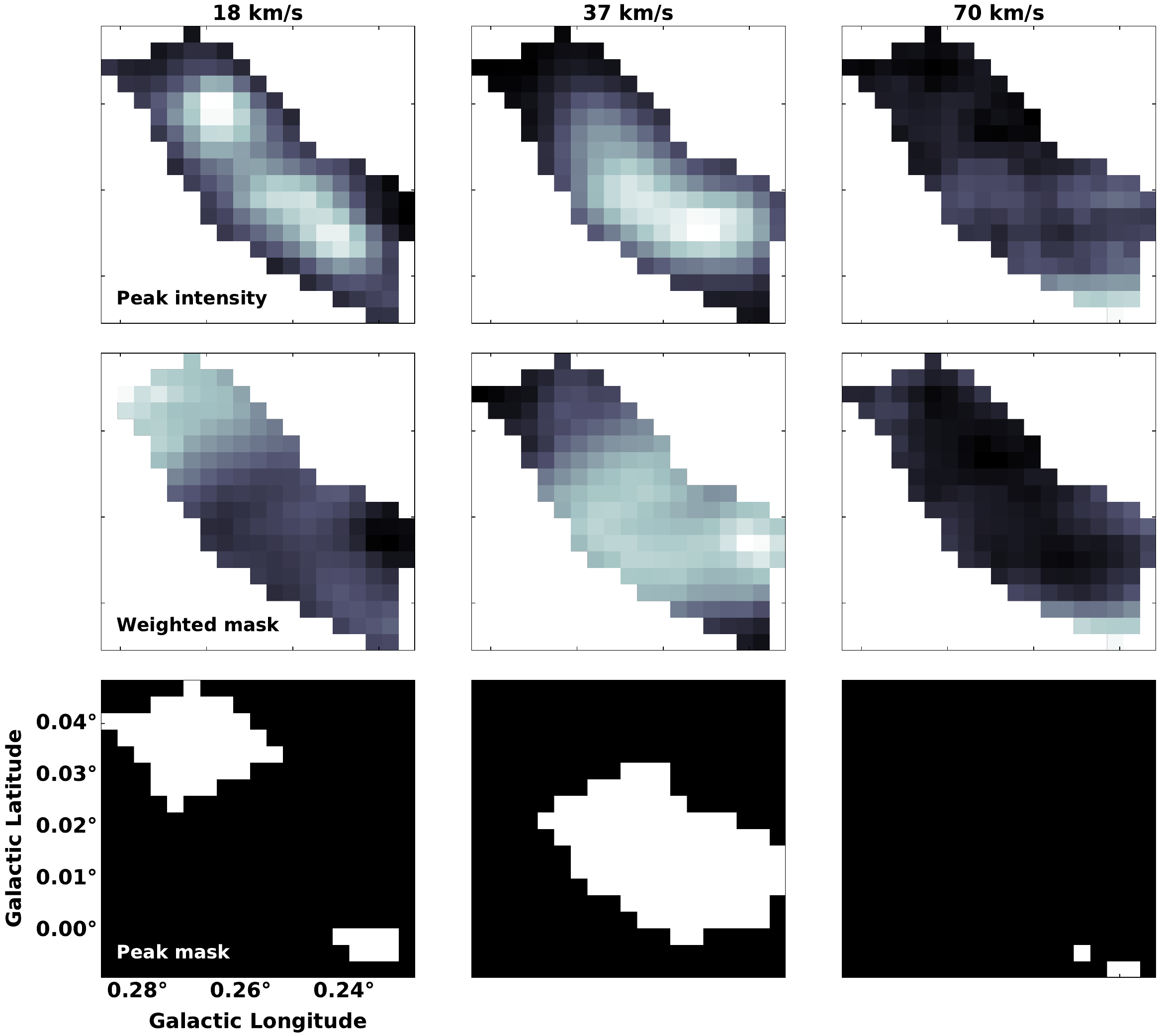}
\caption{An example of the different masking approaches used to account for multiple velocity components when determining their fractional absorption against the radio continuum. The source shown is 17 (\q{the Brick}), which has at least 3 velocity components. \textit{Top row}: peak intensity maps for each of the velocity components. \textit{Centre row}: masks based upon the relative weights of the peak intensities of each component. Each pixel value is a value between 0 and 1, according the the fractional value of the peak intensity of the given component relative to the sum of the peak intensities of all components. \textit{Bottom row}: Traditional binary masks, where for a given pixel, a value of 1 is assigned to the component with the greatest peak intensity value.}
\label{fig:mask_comparison}
\end{figure}

An example of resulting masks are shown in Figure \ref{fig:mask_comparison} for source 17 (\q{the Brick}), which has at least 3 velocity components. The results of the two methods are broadly similar in that they capture the relative 2-D morphological prominence of each component.

Using these masks, we revise our C-band continuum measurements and report two different values. We report the median weighted value, which simply applies the fractional pixel-weighting for each velocity component, as well as the median masked value, which applies the more traditional binary mask per component before measuring the median value of the continuum. 

Having measured both the median continuum emission and maximum spatially-averaged absorption depth for each source, we then take the ratio of the continuum divided by the absorption depth for each of the two masking methods in order to obtain a measure of the fractional absorption. When calculating this ratio we add 2.73~K to the median continuum level to account for the fact that H$_{2}$CO (1$_{1,0}-1_{1,1}$) can be seen in absorption against the CMB. 

Under the assumptions that the absorption is due to the cloud material, and that the continuum emission arises from optically thin free-free or synchrotron emission that fills that Galactic centre at a constant level within \app 100~pc, this fractional absorption gives a simplified diagnostic with which to determine whether a cloud is more likely to be in the foreground or the background of the Galactic centre. If a cloud is in the foreground, then we would expect this ratio to be less than 1 (H$_{2}$CO absorption dominated), whereas if it is in the background we would expect it to be greater than 1 (continuum emission dominated). All measurements for the absorption depth, median continuum, and fractional absorption are given in Table \ref{tab:absorption}.

While this approach works well in many cases, it is fairly crude as it assumes a smooth continuum background. In reality this is not the case, particularly in regions with strong localised continuum emission such as Sgr A* and Sgr B2. Sgr A* in particular is very bright in continuum, and connects to the larger scale, radio-bright Arched filaments. Sgr B2, and to a lesser extent Sgr C, are also very bright due to the embedded high-mass star formation. This means that our approach breaks down in these regions. For example, Sgr B2 has deep correlated absorption across the cloud, but the embedded continuum emission is so bright that the fractional absorption value would place it solidly on the far side of the CMZ, which is inconsistent with the consensus in the literature \citep[e.g.][]{Sawada2004, Reid2009, Chuard2018}. Additionally, the footprint of the ATCA CMZ survey is such that some of Sgr B2 is not covered, though the Sgr B2 Main and North regions are included.

A similar issue is seen for the 20~\kms cloud in that we again see deep, correlated absorption, but the very strong radio continuum from the nearby (in projection) Sgr A* dominates and so our method once again breaks down. This is true for all clouds that are close to Sgr A*. For the 20~\kms cloud, the ATCA map is missing a significant chunk of data right through the centre of the cloud, which means that any analysis using these data needs to be taken with caution. Given the depth of the absorption and how correlated it is in both position and velocity, we conclude that it is more likely that Sgr B2 and the 20~\kms cloud are on the near side of the CMZ.

In the two cases of Sgr B2 and the 20~\kms cloud we use this more qualitative assessment of the absorption depth and position-velocity correlation of the absorption (Section \ref{sec:h2co_absorption}). For the remainder of sources, we use the results from the fractional absorption analysis to determine whether a cloud is more likely to be on the near or far side of the CMZ, or whether the results are inconclusive. Of the clouds/components in our sample, we conclude that 5 are likely are the far side, 18 on the near side, and 8 are uncertain. The primary cause of these uncertainties is projected proximity to Sgr A* and Sgr B2. A discussion of these uncertainties is given in Appendix \ref{ap:abs}. The resulting near/far assignments are given in Table \ref{tab:absorption}.

\begin{deluxetable*}{cccccccccc}
\setlength{\tabcolsep}{3pt}
\tablecaption{Near/far cloud assignments based on H$_{2}$CO (1$_{1,0}$-$1_{1,1}$) absorption and 4.85~GHz continuum emission}
\tablewidth{0pt}
\tablehead{
\colhead{\#} & \colhead{Minimum} & \colhead{Median 4.85 GHz} & \colhead{Fractional} & \colhead{Median 4.85 GHz} & \colhead{Fractional} & \colhead{Velocity} & \colhead{Velocity} & \colhead{Near-side} & \colhead{Colloquial} \\
\colhead{} & \colhead{H$_{2}$CO} & \colhead{continuum} & \colhead{absorption} & \colhead{continuum} & \colhead{absorption} & \colhead{} & \colhead{dispersion} & \colhead{or} & \colhead{name} \\
\colhead{} & \colhead{(1$_{1,0}$-$1_{1,1}$)} & \colhead{(weighted)} & \colhead{(weighted)} & \colhead{(masked)} & \colhead{(masked)} & \colhead{} & \colhead{} & \colhead{Far-side} & \colhead{} \\
\hline
\colhead{} & \colhead{K} & \colhead{K} & \colhead{} & \colhead{K} & \colhead{} & \colhead{km~s$^{-1}$} & \colhead{km~s$^{-1}$} & \colhead{} & \colhead{}
}
\startdata
3 & -2.8 & 6.1 & 2.15 & 6.1 & 2.15 & -90 & 8 & Far & -\\
5 & -11.5 & 6.0 & 0.52 & 6.0 & 0.52 & -78 & 5 & Likely Near & -\\
6a & -1.8 & 4.2 & 2.37 & 5.9 & 3.3 & -29 & 5 & Far & -\\
6b & -2.6 & 4.6 & 1.82 & 5.8 & 2.28 & -21 & 12 & Far & -\\
7a & -18.5 & 5.3 & 0.29 & 6.6 & 0.36 & -73 & 11 & Near & -\\
7b & -3.9 & 4.1 & 1.04 & 7.4 & 1.88 & -37 & 5 & Near & -\\
8a & -5.3 & 6.0 & 1.13 & 12.7 & 2.4 & -54 & 9 & Uncertain & -\\
8b & -0.3 & 11.9 & 35.0 & 11.9 & 34.87 & 15 & 15 & Far & -\\
8c & -4.1 & 6.3 & 1.55 & 11.5 & 2.81 & 62 & 6 & Uncertain & -\\
9 & -8.0 & 21.1 & 3.45 & 21.1 & 2.64 & 15 & 12 & Likely Near$^{*}$ & 20~km/s\\
10 & -97.0 & 75.3 & 0.78 & 75.3 & 0.78 & 48 & 11 & Near$^{*}$ & 50~km/s\\
11a & -9.9 & 17.1 & 1.73 & 27.5 & 2.78 & -11 & 9 & Uncertain$^{*}$ & -\\
11b & 9.0 & 8.8 & 0.97 & 25.2 & 2.8 & 45 & 9 & Uncertain$^{*}$ & -\\
11c & -9.7 & 9.0 & 0.93 & 25.4 & 2.62 & 14 & 8 & Uncertain$^{*}$ & -\\
12 & -6.2 & 24.0 & 3.87 & 24.0 & 3.87 & 86 & 7 & Uncertain$^{*}$ & -\\
13 & -15.3 & 16.2 & 1.06 & 16.2 & 1.06 & 50 & 9 & Likely Near$^{*}$ & Stone\\
14 & -11.4 & 13.0 & 1.14 & 13.0 & 1.14 & 22 & 9 & Likely Near$^{*}$ & Sticks\\
15 & -7.1 & 24.1 & 3.38 & 24.1 & 3.38 & 52 & 11 & Uncertain$^{*}$ & -\\
16a & -23.5 & 9.3 & 0.39 & - & - & -15 & 3 & Near & Straw\\
16b & -21.0 & 13.8 & 0.66 & 18.7 & 0.89 & 57 & 8 & Near & Straw\\
17a & -42.7 & 7.2 & 0.17 & 11.9 & 0.28 & 18 & 16 & Near & Brick\\
17b & -46.7 & 8.7 & 0.19 & 14.3 & 0.31 & 37 & 7 & Near & Brick\\
17c & -9.8 & 3.5 & 0.36 & 16.0 & 1.63 & 70 & 14 & Likely Near & Brick\\
19 & -35.4 & 8.0 & 0.23 & 8.0 & 0.23 & -2 & 13 & Near & -\\
20 & -3.5 & 7.8 & 2.21 & 7.8 & 2.21 & 90 & 18 & Far & Sailfish\\
21a & -19.8 & 5.8 & 0.29 & 7.5 & 0.38 & 8 & 14 & Near & C\\
21b & -7.7 & 4.8 & 0.62 & 8.4 & 1.09 & 39 & 4 & Near & C\\
22 & -45.4 & 6.6 & 0.15 & 6.6 & 0.15 & 19 & 11 & Near & D\\
23 & -22.3 & 7.8 & 0.35 & 7.8 & 0.35 & 28 & 11 & Near & E/F\\
24 & -9.9 & 5.2 & 0.53 & 5.2 & 0.53 & 53 & 12 & Uncertain & -\\
25 & -13.3 & 28.2 & 2.12 & 28.2 & 2.12 & 62 & 12 & Likely Near$^{\dagger}$ & Sgr B2\\\hline\hline
\enddata
\tablecomments{Shown for each velocity component for each source is the absorption depth (minimum of the cloud-averaged ATCA H$_{2}$CO 4.8 GHz line), the median of the C-band continuum emission (GBT, 4.85 GHz) + 2.73~K to account for the absorption against the CMB, the fractional absorption (median C-band continuum divided by the absolute absorption depth), and the velocity and velocity dispersion of each component from the Gaussian fitting of the HNCO emission. The median continuum and fractional absorption values are given twice, corresponding to the different masking techniques discussed in Section \ref{sec:fractional_abs}. This comparison includes 21/31 clouds due to incomplete coverage of the ATCA data. The penultimate column denotes whether a (sub-)cloud is determined to be on the near or far side of the CMZ. Sources marked with an asterisk ($^{*}$) are more uncertain as they are very near to Sgr A* in projection, which complicates this analysis method due to localised strong radio continuum emission. $^{\dagger}$ This is Sagittarius B2, which contains many embedded high-mass star-forming regions, and therefore radio continuum emission. We place Sgr B2 on the near side due to arguments given in Section \ref{sec:fractional_abs}.}
\label{tab:absorption}
\end{deluxetable*}

\section{Discussion}

Having created a CMZ cloud catalogue and, where possible, assigned them to either the near or far side of the CMZ with respect to the Galactic centre, we now compare these results with models of the 3-D orbital geometry. 

\label{sec:discussion}

\subsection{Can the CMZ \textit{really} be described by an ellipse?}
\label{sec:orbit}

The idea that the \app100~pc stream of gas orbiting the Galactic centre can be modelled by a closed ellipse is not a new one. \citet{Binney1991} interpreted the CMZ as roughly elliptical x$_{\textrm{2}}$ orbits. Based on Herschel data, \citet{Molinari2011} proposed a twisted infinity-shaped elliptical orbit that has a radius of \app100~pc and introduces a vertical oscillation to match the projected distribution of the dense gas. While this model does a reasonable job of describing the projected structure of the gas in $l$ and $b$, it has a few notable caveats. Firstly the position of Sgr A* is significantly offset from the geometric centre of the orbit and is placed closer to the near side of the ring (though this is not necessarily problematic, see the discussion later in this subsection). Secondly, the model assumed a constant orbital velocity. Finally, comparison with detailed kinematic analysis of the dense CMZ gas shows that the model is not consistent with the observed structure in position-velocity space \citep{Kruijssen2015, Henshaw_cmz, Henshaw2023}. 

We revisit this model in an effort to address some of these caveats and explore whether a simplified parametric description of an x$_{\textrm{2}}$ orbit can better describe the observed $lbv$ structure of the gas. We initialise an elliptical orbit in Cartesian coordinates as:
\begin{align}
    x &= a \cos(\phi) \\
    y &= -b \sin(\phi) \\
    z &= z_0 \sin(-2\phi + \alpha)
\end{align}
where $a$ and $b$ are the semi-axes, $\phi$ is the azimuthal angle, and $\alpha$ represents a phase offset. The factor of 2 in the sinusoidal term gives rise to the characteristic projected $\infty$ shape in $lb$ space. This form of ellipse is very similar to that in \citet{Molinari2011}, but one crucial difference is that we do not assume a constant orbital velocity, but rather a constant angular momentum. That is, the velocity along the orbit is obtained assuming that the $z$-component of the angular momentum is constant (L$_z$ = $v_{\phi}$R = constant). This is not exactly true either in the case of x$_{\textrm{2}}$-orbits, but is it a much better approximation than a constant orbital velocity.

To fit the model parameters to the data, we initially set the geometric centre of the orbit to be equal to the position of Sgr A* in $l$ and $b$ (-0.0558 deg and -0.0462 deg, respectively). The reasoning is that if an elliptical orbit is related to an x$_{\textrm{2}}$-orbit, then it ought to be centred on the location of the minimum of the global gravitational potential, which is likely Sgr A*, given that the centroid of both the nuclear star cluster and nuclear stellar disc coincide closely with the position of Sgr A* \citep[e.g.][]{Nogueras-Lara2022, Sormani2022}. However, we found that this resulted in a poor fit to the data regardless of other orbital parameter choices. The offset in latitude does match the data well, but the longitudinal offset does not -- the orbit is significantly offset towards negative longitudes compared to the positions of the clouds. We find that an offset equal in magnitude to the longitude of Sgr A*, but in the opposite direction, does a very good job of matching the data. We find that an orbit centred on [$l$, $b$] = [0.05, -0.0462] degrees matches the cloud distribution well (see Figure \ref{fig:lbplot}). 

Though this longitudinal offset is an explicit by-eye match to the data, the offset may be explained by various expected processes. In particular, numerical simulations show that gas accretion on to the CMZ via bar-driven inflow is unsteady, which can significantly perturb the gas \citep{Sormani2018b}, and has been proposed to explain the highly asymmetric longitudinal dense gas distribution observed in the CMZ (as visible in Figure \ref{fig:colmap}). Furthermore, simulations show that while the structure of the CMZ gas is well described by a ring-like structure when time-averaged, it is highly variable as a function of time, with substantial asymmetries and irregular structures when observed at any individual snapshot (see e.g.\ Figure 6 of \citealt{Tress2020}). Given that we can only observe the real CMZ in a single discrete snapshot, it is reasonable to expect that it may be irregular and asymmetric. Simulations presented in \citet{Emsellem2015} also demonstrate a decoupling of Sgr A* from the circumnuclear gas ring, with offsets of up to 70~pc. In summary, we do not expect that the CMZ should look like a simple ring centred on Sgr A* at any given instantaneous snapshot, and so the longitudinal offset from Sgr A* of \app 0.09 deg (13~pc in projection at 8.1~kpc) for the elliptical model presented here is minor and could easily be explained by the turbulent CMZ gas dynamics.

\begin{figure*}
\includegraphics[width=\textwidth]{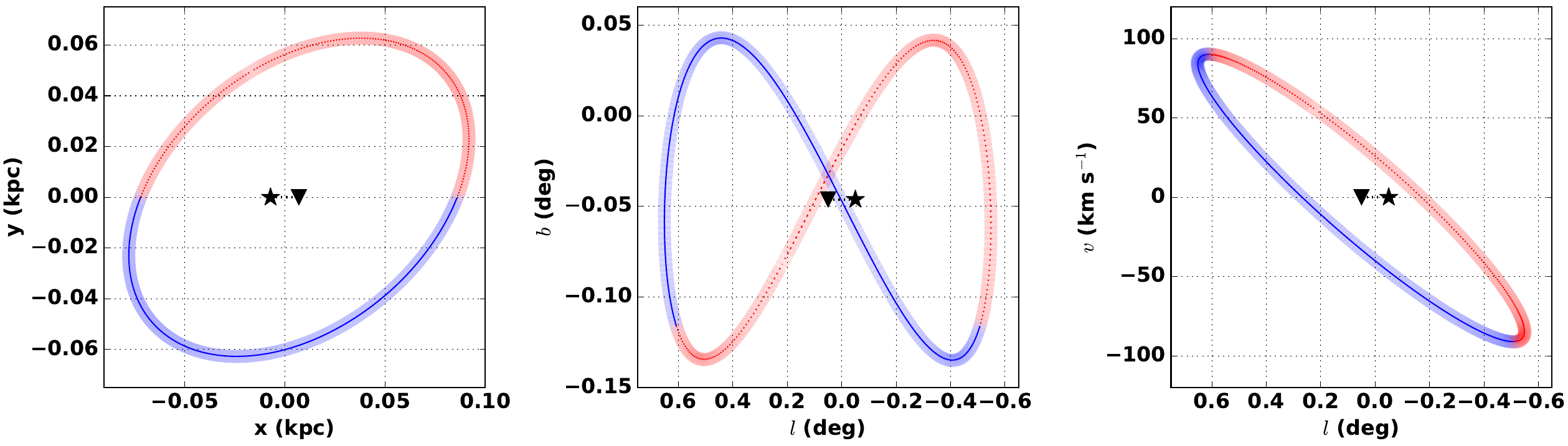}
\caption{Geometric toy model of a 3-D vertically-oscillating elliptical orbit. \textit{Left}: A top-down view of the orbit in Cartesian coordinates. \textit{Centre}: Galactic longitude vs. latitude. \textit{Right}: Galactic longitude vs. velocity. The blue solid lines correspond to the segments of the model that are in front of the Galactic centre, and red dotted lines to those behind the Galactic centre. The star marker denotes the location of Sgr A*, and the inverted triangle shows the  centre of the orbit.}
\label{fig:elliptical_orbit}
\end{figure*}

Given these central coordinates, we find that the following parameters fit the data reasonably well: $a$ = 90~pc, $b$ = 55~pc, $z_{0}$ = 12.5~pc, and $v_0$ = 130~\kms, where $z_{0}$ is the height of the orbit, and $v_0$ is the initial tangential velocity. We also add a rotation about the $z$-axis (i.e. the angle between the orbital major axis and the Sun-Galactic centre line), $\theta$. We find that $\theta$ = 25 deg provides the best match to the data. The resulting orbit\footnote{The code used to generate this toy model is available at: \url{https://github.com/CentralMolecularZone/3D_CMZ}} is shown in Figure \ref{fig:elliptical_orbit}. 

Note that the model parameters were chosen through a by-eye process. The motivation here is simply to investigate whether a closed elliptical orbit can do a good job of describing the $lbv$ structure of the dense gas. A thorough exploration of the parameter space and proper physical modelling is deferred to a subsequent paper in this series.

\subsection{Qualitative comparison with 3-D models of CMZ geometry}
\label{sec:3dcmz}

Using the near/far assignments of clouds determined in Section \ref{sec:fractional_abs}, we can directly compare these with existing models of the 3-D structure of the dense gas in the CMZ. As discussed in Section \ref{sec:intro}, there are broadly three main flavours of geometrical models that are commonly discussed in the literature. To recap, these models interpret the gas structure as (i) two spiral arms \citep[e.g.][]{Sofue1995a, Ridley2017}, (ii) a vertically-oscillating, closed ellipse \citep[e.g.][and the updated model described in Section \ref{sec:orbit}]{Molinari2011, Sofue2022}, and (iii) an open, eccentric orbit that appears as a pretzel-shaped collection of gas streams in projection \citep{Kruijssen2015}.

\begin{figure*}
\includegraphics[width=\textwidth]{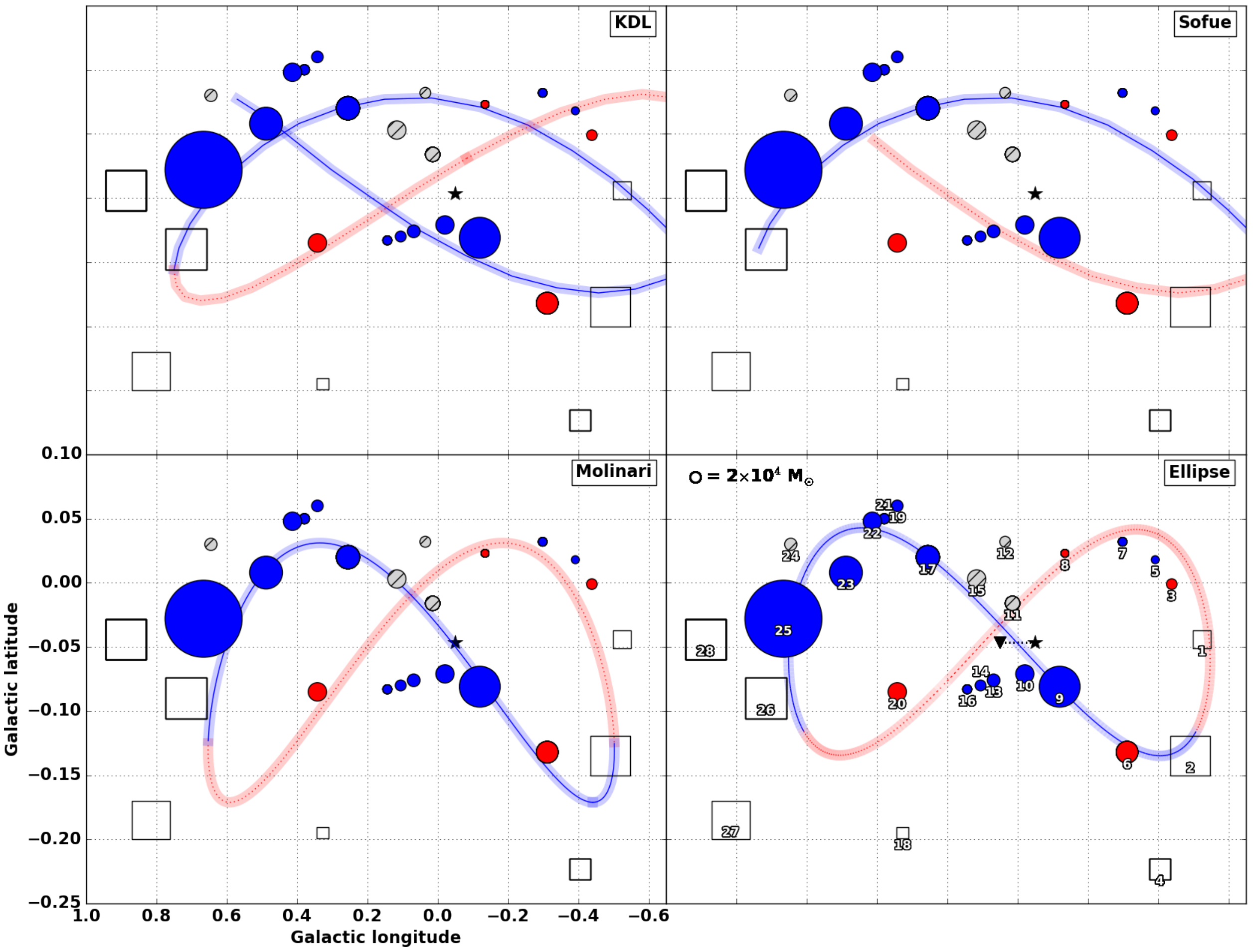}
\caption{Coloured lines show the 2D projection of four separate orbital models in position-position (\textit{top left}: \citet{Kruijssen2015} [KDL], \textit{top right}: \citet{Sofue1995b} [Sofue], \textit{bottom left}: \citet{Molinari2011} [Molinari], \textit{bottom right}: a vertically-oscillating elliptical orbit, similar to that of \citet{Molinari2011}) [Ellipse]. Note that the spiral arms model from \citet{Sofue1995a} is not explicitly presented in their paper -- the visualisation shown here is adapted from two streams of the \citet{Kruijssen2015} model, as presented in \citet{Henshaw2016b}. The blue solid lines correspond to the regions of the orbital models that are in front of the Galactic centre, and red dotted lines to those behind the Galactic centre. Markers correspond to the positions of the clouds in the catalogue presented in this paper, the sizes of which are related to the cloud mass. Coloured circles show clouds for which both ATCA H$_{2}$CO (1$_{1,0}$-$1_{1,1}$) data and GBT 4.8 GHz continuum data are available. The marker colours correspond to the presence of deep (\textgreater\ 2.73~K), absorption at the velocity of the cloud emission which is correlated in position and velocity. Blue markers correspond to the presence of such absorption and are concluded to more likely be in the foreground, whereas red markers do not show such absorption and are more likely to be in the background relative to the Galactic centre. Grey hatched markers are uncertain  due to low signal-to-noise, and/or artifacts in the ATCA data. Open black square markers correspond to clouds in the catalogue for which these data are not available. In the bottom right panel, the structure ID for each source is given alongside the corresponding marker. The star marker denotes the location of Sgr A*, and in the lower right panel the inverted triangle shows the centre of the orbit.}
\label{fig:lbplot}
\end{figure*}

In the following, we compare our cloud near/far assignments with four specific versions of these models: spiral arms \citep{Sofue1995a}, closed ellipse with constant orbital velocity \citep{Molinari2011}, closed ellipse with constant angular momentum (this paper), and the open stream model \citep{Kruijssen2015}. The $lb$ projections of these four models are shown in Figure \ref{fig:lbplot}, and are labelled as \q{Sofue}, \q{Molinari}, \q{Ellipse}, and \q{KDL}, respectively. Note that \citet{Sofue1995a} do not provide an explicit model for their spiral arm interpretation. The \q{Sofue} model shown in all plots in this paper is an approximation of the spiral arms model that is based upon the \citet{Kruijssen2015} model \citep{Henshaw2016b}.

In Figure \ref{fig:lbplot} (and all subsequent figures), each model is divided into near and far components. Red dotted segments of the orbit are those that are behind Sgr A*, and blue solid segments are in front of Sgr A*. Overlaid on each model are the clouds from the catalogue presented in this paper, excluding those beyond $l$ \app1~deg, as (i) they are likely not part of the circumnuclear stream, and (ii) we do not have absorption data there. Circular markers show the 21 clouds for which we have absorption data, the sizes of which correspond to the cloud mass. The colour of the markers corresponds to the near or far assignment of the cloud, where blue indicates absorption-dominated (i.e. near-side) and red indicates emission-dominated (i.e far side). Grey markers are uncertain due to low signal-to-noise, and confusion or artifacts in the ATCA data from Sgr A* and Sgr B2 (see Appendix \ref{ap:abs}).

We see that all four models generally do a reasonable job of describing the $lb$ distribution of the clouds, which is to be expected as this is how they were defined. Folding in the near/far assignments of the clouds, we begin to see clearer differences between the models. Overall, the KDL model provides to best match to our results in $lb$-space. The spiral arms model shows the least consistency with our cloud positions, which places the so-called \q{Three Little Pigs} clouds \citep[IDs 13, 14, 16;][]{Battersby2020}, and the 20 and 50~\kms clouds (IDs 9 and 10) on the far side. We find that these clouds all show significant absorption, which we interpret as them being more likely to be in front of the Galactic centre. However, this region is notoriously difficult to interpret, particularly the 20 and 50~\kms clouds, as the clouds reside very close to Sgr A* in projection \citep[a detailed review can be found in][]{Henshaw2023}. In our approach, this proximity complicates the analysis due to the very bright radio continuum emission, and artifacts in the ATCA line data (Section \ref{sec:fractional_abs}.) 

One notable outlier with the two elliptical models and the open stream model is source 6, which lies in the lower right quadrant of the orbit. Our analyses tentatively place this cloud on the far side of the CMZ. There is absorption at the cloud velocity, but it is very weak (Table \ref{tab:absorption}). If this interpretation is correct, then this is inconsistent with the elliptical and open stream models, and it is consistent with the spiral arms model. In a separate paper in this series, Lipman et al. (submitted, \textit{Paper IV}), using independent methodologies based on IR extinction, are unable to confidently assign a near/far placement for this cloud, though their results tentatively suggest that the lower velocity component (6a) is likely on the near side. This highlights the requirement for independent, multi-wavelength studies of these clouds. In general they find good agreement between the near/far placements of clouds between their methods and ours, but there are some outliers, such as this one. We refer the reader to \textit{Paper IV} for a detailed comparison between their results and those presented in this paper.

\begin{figure*}
\includegraphics[width=\textwidth]{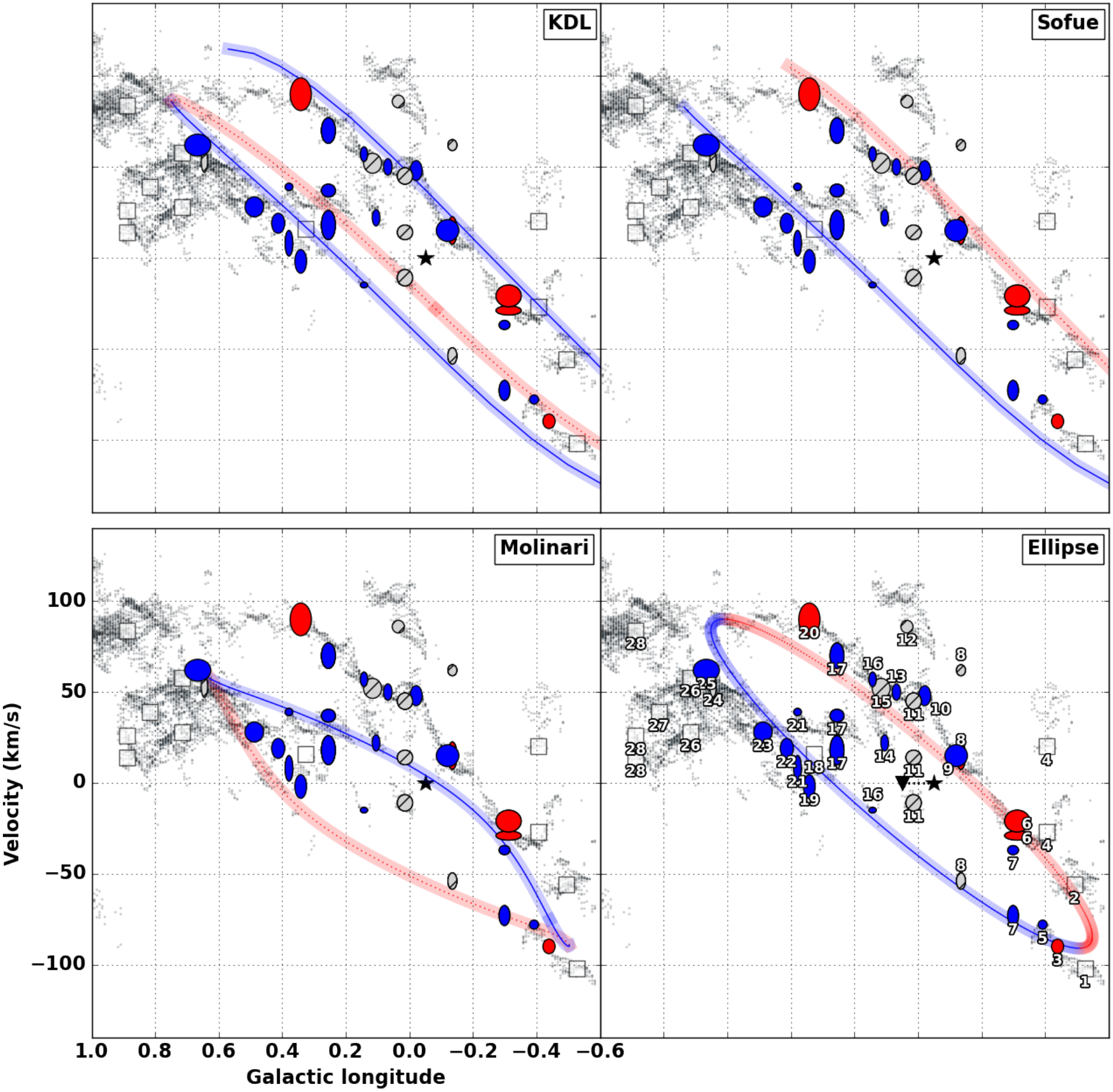}
\caption{Coloured lines show the 2D projection of four separate orbital models in longitude-velocity (\textit{top left}: \citet{Kruijssen2015} [KDL], \textit{top right}: \citet{Sofue1995b} [Sofue], \textit{bottom left}: \citet{Molinari2011} [Molinari], \textit{bottom right}: a vertically-oscillating elliptical orbit, similar to that of \citet{Molinari2011} [Ellipse], though they assume constant orbital velocity, whereas we assume constant angular momentum). As in Figure \ref{fig:lbplot}, the solid blue lines correspond to the regions of the orbital models that are in front of the Galactic centre, and dotted red lines to those behind the Galactic centre. Markers correspond to the positions of the clouds in the catalogue presented in this paper. The sizes of the markers in $l$ and $v$ correspond to their angular radius and velocity dispersion, respectively. Coloured circles show clouds for which both ATCA H$_{2}$CO (1$_{1,0}$-$1_{1,1}$) data and GBT 4.8 GHz continuum data are available. The marker colours correspond to the presence of deep (\textgreater\ 2.73~K), absorption at the velocity of the cloud emission which is correlated in position and velocity. Blue markers correspond to the presence of such absorption and are concluded to more likely be in the foreground, whereas red markers do not show such absorption and are more likely to be in the background relative to the Galactic centre. Grey hatched markers are uncertain  due to low signal-to-noise, and/or artifacts in the ATCA data. In the case of multiple velocity components, each pixel in the leaf for a given component is assigned a weight according to the fractional peak intensity relative to the integrated peak intensity across all components. When evaluating the fractional absorption for each velocity component, these masks are applied to the radio continuum maps in order to apply a relative weight to each component based on the morphological distribution of the peak intensity (Section \ref{sec:fractional_abs}). Square markers correspond to clouds in the catalogue for which the relevant data are not available. The background grey points correspond to the spectral decomposition of MOPRA HNCO data as presented in \citep{Henshaw2016b}. The star marker denotes the location of Sgr A*, and in the lower right panel the inverted triangle shows the centre of the orbit.}
\label{fig:lvplot_weightmask}
\end{figure*}

Figure \ref{fig:lvplot_weightmask} shows the same model vs. cloud comparison, this time in $lv$-space. Here the cloud markers are scaled independently along both axes, corresponding to the effective radius and velocity dispersion of each. In this projection, the differences between the models are more pronounced. In particular the \citet{Molinari2011} elliptical model provides a very poor fit the $lv$ data, and captures very little of the higher-velocity emission \citep[see also][]{Kruijssen2015, Henshaw2016b, Henshaw2023}. In contrast, the similar elliptical model presented in Section \ref{sec:orbit} is more consistent with the data. The primary reason for this difference is the assumption of constant orbital velocity in the \citet{Molinari2011} model, whereas we assume constant angular momentum. We therefore conclude that while the \citet{Molinari2011} model describes the $lb$ structure well, it is not a suitable model based on kinematic arguments.

The updated ellipse, spiral arms, and open stream models each provide a good fit to the majority of the observed $lv$ structure, though there are certain regions that are not well described by any models, such as the extended material beyond Sgr B2 ($l$ $\gtrsim$ 0.65 deg), and the central region of the $lv$ plot at $l$ \app 0.0 -- 0.4 degrees, and $v$ \app 20 -- 40~\kms. We discuss these poorly fitted regions in more detail in \ref{sec:other_lv}. All three models place the dust ridge (the clouds between $l$ \app0.2 -- 0.5~deg and $v$ \app0 -- 40~\kms) and Sgr B2 on the near side, which is consistent with other observational constraints on these regions \citep[e.g.][]{Reid2009, Chuard2018, Nogueras-Lara2021}.

Disagreements between these three models are more apparent when looking at the near/far positions of the clouds. The updated ellipse and spiral arm models are broadly consistent with each other in that they place the lower velocity segments of the streams in the foreground, and the higher velocity segments in the background. The lower velocity near-side segments show good agreement with our near/far assignments, with the exception of sources 6 and 8, which we place on the far side based on the lack of absorption. This is potentially resolved by the \citet{Kruijssen2015} model, where the far-side stream is very close to the lower velocity near-side stream in $lv$-space.

The most significant discrepancies can be seen in the higher velocity portion of the orbits. The updated ellipse and spiral arms models place this material on the far side, whereas the open stream model places it all on the near side. Placing this gas on the near side is consistent with the fact that it contains the \q{Three Little Pigs} and 20 \& 50~\kms clouds, which, as discussed earlier, we conclude are in the foreground based on deep absorption. However, as can be seen in Figure \ref{fig:lvplot_weightmask}, our analyses place some of the clouds belonging to this high velocity gas on the far side also. This mix of near and far side clouds in the higher velocity segment is not consistent with any of the current orbital models. Overall, we see that the KDL model provides the best match to our near/far assignments in $lv$-space, the ellipse and Sofue models have more near/far mis-classifications, particularly in the higher velocity gas, and the Molinari model shows the least consistency in both $lbv$ structure and near/far assignments.

In summary, we find that all of the models have some level of inconsistency with the data. The \citet{Molinari2011} model is ruled out based on kinematic arguments. The spiral arm model from \citet{Sofue1995a} and the toy elliptical model presented in this work describe the $lbv$ structure reasonably well, but there are inconsistencies with the placement of clouds in the Sgr A region, some of which are placed on the far side, in conflict with our radio absorption analysis which places them on the near side. The \citet{Kruijssen2015} model is the most consistent in $lbv$-space, and does a good job of matching both the observed structure and many of the near/far placements. However, there are still some inconsistencies, particularly in their higher velocity near-side stream, in which we find clouds that are concluded to be on the far side based on our analyses in this paper. Furthermore, an open stream as presented in \citet{Kruijssen2015} has not been successfully reproduced in simulations \citep{Armillotta2020, Tress2020}.

\subsection{Quantitative comparison with 3-D models of CMZ geometry}
\label{subsec:quant_comparison}
To provide a more quantitative comparison between the different orbital models, we use several metrics to assess how well each model matches the observational data in $lbv$ space and in terms of near/far classification at the locations of the clouds in our catalogue. 

We first preprocess the data by resampling all four orbital models on to a common grid of $lbv$ points, each with a near/far classification. The near and far classifications are mapped to 0 and 1, respectively. The model points and the data from our catalogue in Table \ref{tab:absorption} are then standardised in $lbv$ space using the \texttt{RobustScaler} from \texttt{scikit-learn}. This method centres the data by subtracting the median and scales it according to the interquartile range (IQR), reducing the influence of significant outliers compared to methods that use mean and standard deviation. However, while \texttt{RobustScaler} adjusts for outliers and brings each feature onto a comparable scale, it does not address differences in units between the dimensions or account for correlations between them.

For the $lbv$ comparison, we calculate the overall distance between each model and the data in 3-D using the Mahalanobis distance metric via the \texttt{cdist} function in \texttt{scipy}. The Mahalanobis distance metric accounts for the covariance structure of the data, effectively handling the different units and correlations between the $lbv$ dimensions. This provides a more accurate measure of similarity in $lbv$ space compared to a standard Euclidean distance. The overall distance between the data and the model is then computed as the average minimum normalised $lbv$-distance, on a scale of 0 to 1. 

\begin{figure*}
\centering
\includegraphics[width=\textwidth]{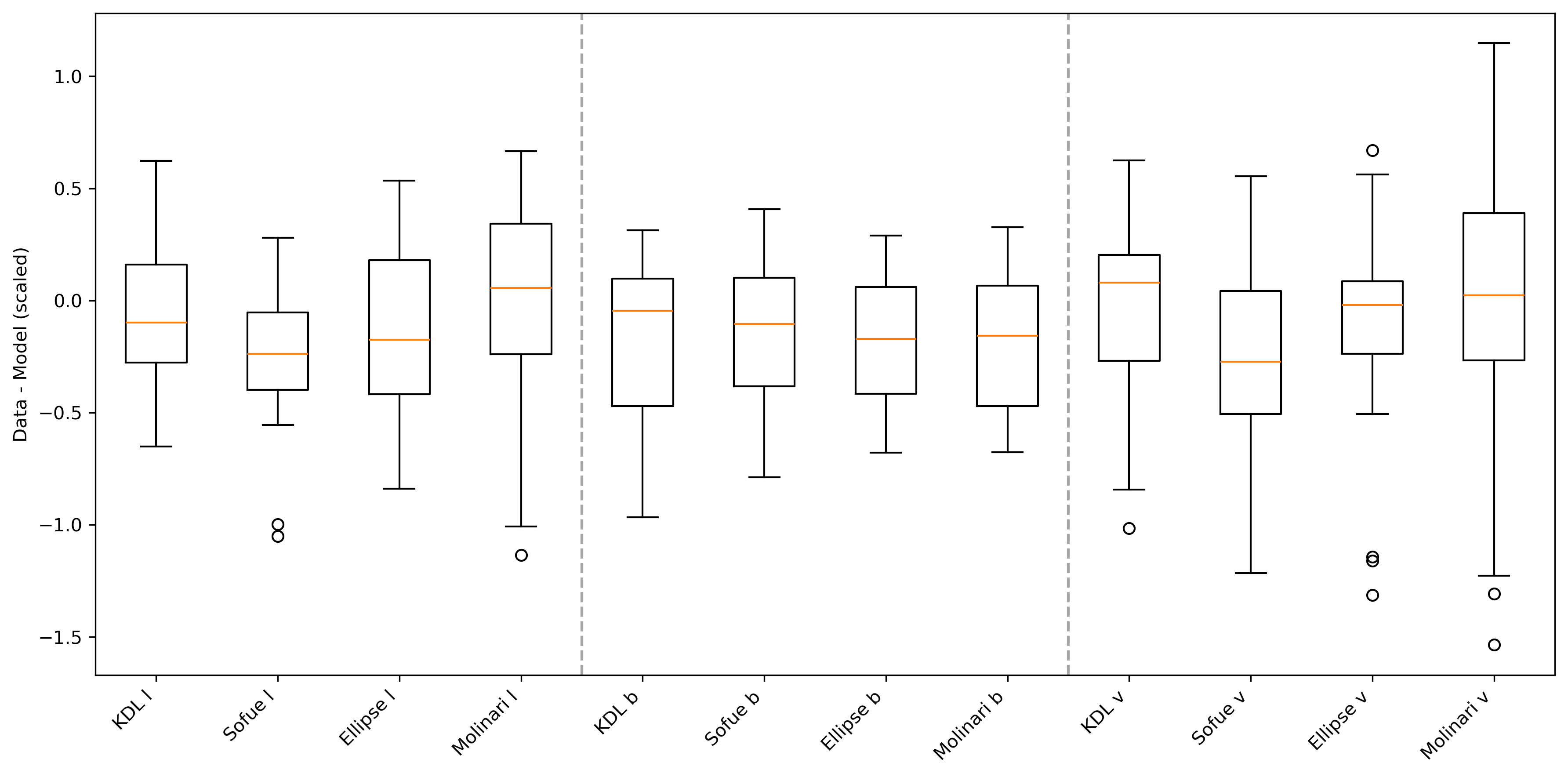}
\caption{Distribution of differences between the scaled observed cloud positions and the nearest model points for each orbital model in longitude ($l$), latitude ($b$), and velocity ($v$).}
\label{fig:error_distribution}
\end{figure*}

Figure~\ref{fig:error_distribution} shows the distribution of the scaled difference between the cloud catalogue and the model points for each dimension in $lbv$. We see that all models tend to perform worse in velocity space, especially the Molinari model, likely due to the complex kinematics in the CMZ.

To assess the near/far predictions, we use a k-nearest neighbours (kNN) approach via \texttt{scitkit-learn}'s \texttt{NearestNeighbors} functionality. For each cloud, we find the k-nearest model points in $lbv$ space and take a weighted vote of their near/far predictions, where the weight of each neighbour is based on the inverse of the distance between it and the data, effectively giving more weight to the closest neighbours. The choice of k for the number of neighbours is set to $\sqrt{N}$, where $N$ is the number of points in the model. This is a typical \q{rule of thumb} for a good starting point, and we confirm that the following results are robust up to at least k $\approx$ n/4, where n is the number of model points. We finally compute the near/far accuracy (on a scale of 0 to 1) as the weighted sum of correct predictions divided by the total sum of weights.

We find that the KDL model performs best overall, with good $lbv$ matching (distance of 0.22) and near/far accuracy (71\%). The Sofue model shows a slightly better $lbv$ match (0.19) but lower near/far accuracy (64\%). The ellipse model presented in this work ranks third, with the largest $lbv$-distance (0.28) and reasonable near/far accuracy (55\%). The Molinari model performs poorest overall, particularly in near/far predictions (37\%), and with the second largest $lbv$-distance (0.25). The results of these analyses are summarised in Table~\ref{tab:model_comparison}. 

We find that the absolute near/far accuracy scores vary as a function of k in kNN, but the relative ranking of the model scores is unchanged. Beyond a k-value of n/4, the number of neighbours begins to approach a significant fraction of the total sample size, and the resulting scores tend to become noisier and less stable as k increases. We also test the sensitivity to the strength of the distance-based weighting. The model accuracy shown in the second column of Table~\ref{tab:model_comparison} weights each neighbour as 1/$d^{\alpha}$, where $d$ is the distance between the data and the neighbouring model point and $\alpha$ = 1. As we increase this from $\alpha$ = 1 $\rightarrow$ 2, we find that the accuracy scores for the Sofue and Ellipse model begin to converge, and eventually flip such that the Ellipse model performs better when giving more weight to closer neighbours. This trend is robust to further increases in $\alpha$. We therefore also report the scores for $\alpha$ = 2 in Table~\ref{tab:model_comparison}.

\begin{table}
\centering
\caption{Quantitative comparison of orbital models. The near/far accuracy shows the percentage of clouds in our catalogue are correctly classified as near or far side for each model. The two separate columns for this show the different accuracy scores corresponding to the strength of the distance-based weighting in the k-nearest neighbour analysis, where $\alpha$ is the exponent in the 1/$d^{\alpha}$ weighting scheme. Shown also is the agreement between the model and the data using the method from \textit{Paper IV}, as summarised in section \ref{subsec:quant_comparison}. The $lbv$-distance is the average minimum of the normalised distances between the data and the model.}
\label{tab:model_comparison}
\begin{tabular}{lcccc}
\hline
Model & Near/far & Near/far & \textit{Paper IV} & $lbv$ \\
& accuracy & accuracy & method & distance \\
& $\alpha$ = 1 & $\alpha$ = 2 & & \\
\hline
KDL & 74\% & 80\% & 64\%  &0.22 \\
Sofue & 62\% & 59\% & 64\% & 0.19 \\
Ellipse & 58\% & 67\% & 55\% & 0.28 \\
Molinari & 36\% & 34\% & 41\% & 0.25 \\
\hline
\end{tabular}
\end{table}

An independent method for quantifying the comparison between the models and the data is presented in \textit{Paper IV}. We refer the reader to Lipman et al. (submitted) for a detailed explanation of this method. In brief, the $lbv$ distance calculation is performed similarly to the method presented here, instead using min-max scaling and a Euclidean distance metric. To assess the near/far classification accuracy, the authors instead impose a distance threshold (0.25 in normalised \textit{lbv} space), where the minimum distance between the model and the data must be within this threshold in order to be considered a positional match. If a source is a match in $lbv$ space, then the near/far classification of the data and model are compared, and if they agree, then this is added to the accuracy score for that model. The result of applying this method to the data presented in this paper is also shown in Table~\ref{tab:model_comparison}. Overall we find that the accuracy scores and relative model ranking are comparable to those using the kNN method.

These results support our earlier qualitative assessment that the KDL model provides the best overall match to the observational data, and most accurately predicts the near/far classification of the clouds. However, they also highlight that significant discrepancies remain between all current models and the observed CMZ structure.

The poor performance of the Molinari model, particularly in near/far predictions, quantitatively confirms our earlier conclusion that it is not a suitable description of the CMZ structure. While our updated ellipse model shows improvement, especially in near/far accuracy, its relatively poor $lbv$ matching suggests that this simple closed elliptical orbit may be too restrictive to fully describe the observed gas distribution, or that our parameter choices for the model are not well optimised. More in-depth modelling and statistical analyses of such an orbit will be explored in a subsequent paper in this series.

Overall, these analyses reinforce the need for more complex models of the CMZ's 3-D structure. While the open stream configuration of the KDL model appears to best match the data presented in this work, it does not fully explain the underlying gas distribution. Obtaining more precise observational constraints on cloud distances will be crucial for discriminating between competing models of the CMZ's 3-D geometry, and developing improved ones.

\subsection{What are the origins of the poorly-constrained structures at $l$ \textgreater\ 0.7$^{\circ}$}
\label{sec:other_lv}

As noted in the previous section, none of the existing models of the CMZ's 3-D structure provide an exact match to the data. This is true in $lbv$-space generally, but it is even more pronounced in $lv$-space.

One region of the CMZ that is not well described by any of the models in $lbv$-space is the material beyond Sgr B2, roughly between 0.7$^{\circ}$ \textless\ $l$ \textless\ 1.1$^{\circ}$. The bulk of the gas in this region, which contains our sources 27, 28, and 29, falls outside of the $lbv$ extents of all of the models discussed here. Sources 30 and 31 are also not captured by any of the models, but they are much further out from the CMZ at \app 1.6$^{\circ}$, and so we do not include these in the context of 3-D CMZ models.

A possible explanation for this 0.7$^{\circ}$ \textless\ $l$ \textless\ 1.1$^{\circ}$ region it that is related to material that is flowing into the CMZ from the bar. It is well established that the Galactic bar drives mass into the CMZ along the dust lanes \citep[e.g.][]{Sormani2019b}. The so-called 1.3$^{\circ}$ complex, which is connected to this 0.7$^{\circ}$ \textless\ $l$ \textless\ 1.1$^{\circ}$ region in velocity, is considered to be a strong candidate for infalling material that is directly interacting with the main CMZ ring-like structure, as evidenced by extended velocity features \citep[e.g.][]{Sormani2019a, Tress2020}. Given that these regions are connected in both position and velocity, it is plausible that this material is also part of the interaction between the CMZ and the inflowing gas, and is therefore unlikely to be well described by a simple geometric model of the circumnuclear orbit.

Another possibility is that the material at \app 1$^{\circ}$ is overshooting the CMZ. Simulations demonstrate that a significant fraction (up to 70~\%) of the material accreted via the dust lanes on one side of the CMZ may overshoot the main orbit, and then be accreted on to the main orbit at the opposite side \citep{Sormani2019a, Hatchfield2021}.

\subsection{The asymmetric CMZ}

It has long been noted that the projected distribution of dense gas in the CMZ is highly asymmetric. The majority of the dense gas and molecular clouds are at positive longitudes, and mostly concentrated in the dust ridge and the Sgr B2 complex, extending out to the 1.3$^{\circ}$ complex and beyond \citep[see Figure \ref{fig:colmap}, e.g.][]{Bally1988, Molinari2011, Longmore2013b}. 

While this longitudinal asymmetry is well established, the distribution of molecular clouds in front of vs.\ behind the Galactic centre is not known. In other words, do we see any similar asymmetries in the dense gas along the line of sight? Though we do not have measurements of the line of sight distances of the clouds in this work, we do now have the first complete catalogue of CMZ clouds with constraints on their relative positions, and so we can begin to explore this question.

As can be seen in Figure \ref{fig:lbplot} the majority of the clouds in our catalogue are more consistent with being in front of the Galactic centre, with 13 clouds in the foreground, 4 in the background, and 4 uncertain. This changes to 18 foreground, 5 background, and 8 uncertain when accounting for multiple velocity components (see Figure \ref{fig:lvplot_weightmask} and Table \ref{tab:absorption}). Of the 18 in the foreground, 13 of these are at 0$^{\circ}$ \textless\ $l$ \textless\ 0.7$^{\circ}$. In summary, of the 23 sources with near/far assignments, 78~\% are on the near side, and 57~\% are at $l$ \textgreater\ 0$^{\circ}$. The CMZ molecular cloud distribution is strongly asymmetric along the line of sight, with the majority of clouds residing in front of the Galactic centre, most of which are at positive longitudes.

As discussed in Section \ref{sec:orbit}, such asymmetries are not necessarily unexpected, as they are commonly seen in simulations as a result of the unsteady flow of material along the dust lanes towards the CMZ \citep[e.g.][]{Sormani2018b}. Combined with the effects of feedback and overshooting material, it is common to see asymmetries and highly perturbed structure in such simulations \citep{Tress2020}. Strong asymmetries are also seen in observations of extragalactic CMZs, such as the inner circumnuclear ring in M83, where as much as two thirds of the material are contained on one side of the ring \citep{Callanan2021}.

\section{Conclusions}
\label{sec:conclusions}

We have presented a catalogue of clouds in the CMZ along with their global physical and kinematic properties (Table \ref{tab:leafgeneralproperties}). We obtain new constraints on the line-of-sight positions of 21 CMZ clouds relative to the Galactic centre using radio molecular line absorption analysis (Section \ref{sec:h2co_absorption}, Table \ref{tab:absorption}). We have also introduced an updated version of a geometric model describing the CMZ gas as a closed, vertically-oscillating eccentric orbit (Section \ref{sec:orbit}).

Using these new constraints on which clouds are in front of vs. behind the Galactic centre, we compare our results with four different models of the 3-D geometry of the CMZ gas. We find that no single model in the literature adequately fits our results. The open stream model from \citet{Kruijssen2015} shows the most consistency with the data, though there are still discrepancies between the near/far placement of several clouds, particularly in the higher velocity gas stream, and the physical origin of such an orbit is not clear.

We also find that the distribution of molecular clouds in the CMZ is highly asymmetric along the line of sight, with the majority of clouds residing in front of the Galactic centre, and at positive longitudes.

These results highlight that the CMZ is likely significantly more complex than can be captured by relatively simple geometric models \citep[e.g.][]{Tress2020}, and that more observational data is needed to provide stronger constraints on the line-of-sight positions of CMZ clouds. In the context of this paper, targeted high angular resolution molecular line absorption data would facilitate much more precise analyses of the substructure of the absorbing gas. In the context of the wider 3-D CMZ project, the use of more novel, multi-wavelength techniques, such as maser proper motion measurements, dust extinction methods using JWST data, and x-ray echoes through molecular clouds from past Sgr A* flaring events, all promise to provide new constraints on the line-of-sight positions of molecular clouds, and ultimately the true present-day 3-D geometry of the dense CMZ gas.

\section*{Acknowledgements}

D. L. Walker, C.\ Battersby, and D. Lipman gratefully acknowledge support from the National Science Foundation under Award No. 1816715. 
C.\ Battersby  also gratefully  acknowledges  funding  from  National  Science  Foundation (NSF)  under  Award  Nos. 2108938, 2206510, and CAREER 2145689, as well as from the National Aeronautics and Space Administration through the Astrophysics Data Analysis Program under Award ``3-D MC: Mapping Circumnuclear Molecular Clouds from X-ray to Radio,” Grant No. 80NSSC22K1125. 
D. Lipman also gratefully acknowledges funding from the National Science Foundation under Award Nos. 2108938, and CAREER 2145689, as well as from the NASA Connecticut Space Grant Consortium under PTE Federal Award No: 80NSSC20M0129.
M. C. Sormani acknowledges financial support from the European Research Council under the ERC Starting Grant ``GalFlow'' (grant 101116226).
A. Ginsburg acknowledges funding from NSF awards AST 2008101 and 2206511 and CAREER 2142300.
J.\ D.\ Henshaw gratefully acknowledges financial support from the Royal Society (University Research Fellowship; URF/R1/221620). 
R. S. Klessen and S. C. O. Glover acknowledge financial support from the European Research Council via the ERC Synergy Grant ``ECOGAL'' (project ID 855130),  from the German Excellence Strategy via the Heidelberg Cluster of Excellence (EXC 2181 - 390900948) ``STRUCTURES'', and from the German Ministry for Economic Affairs and Climate Action in project ``MAINN'' (funding ID 50OO2206). 
E. A. C. Mills  gratefully  acknowledges  funding  from the National  Science  Foundation  under  Award  Nos. 1813765, 2115428, 2206509, and CAREER 2339670. 
Q. Zhang acknowledges the support from the National Science Foundation under Award No. 2206512.

\section*{Data Availability}

The relevant data products and code used for this paper and other papers in this series are made available at \url{https://centralmolecularzone.github.io/3D_CMZ/}

An interactive version of the 3-D models and cloud catalogue presented in this paper are made available at \url{https://3d-cmz-models.streamlit.app/}

\section*{software}

This research primarily made use of the following software packages: \textbf{CASA} \citep{casa}, \textbf{astropy}, a community-developed core Python package for Astronomy \citep{astropy:2013, astropy:2018}, \textbf{astrodendro}, a Python package to compute dendrograms of astronomical data (\url{http://www.dendrograms.org/}), \textbf{APLpy}, an open-source plotting package for Python \citep{aplpy}, \textbf{spectral-cube} (\url{https://spectral-cube.readthedocs.io/en/latest/}), \textbf{radio-beam} (\url{https://radio-beam.readthedocs.io/en/latest/}).

\appendix
\counterwithin{figure}{section}
\section{Multi-component Gaussian fitting}
\label{ap:gauss}

As outlined in Section \ref{sec:kinematic_properties}, we select both HNCO and H$_{2}$CO as our primary lines to probe the global kinematic properties of the clouds in our catalogue. For the purpose of this paper, we are interested only in the general kinematics, i.e. the centroid velocity and velocity dispersion, as we are simply comparing with the PPV structure of various models of 3-D CMZ structure. Thorough decomposition of the spectra is beyond the scope of this work, and has already been presented for the full CMZ using HNCO in \citet{Henshaw2016b}. We therefore opt to fit Gaussian components to each cloud-averaged spectrum to obtain these properties. To do this we use \texttt{pyspeckit}, and for each cloud we manually inspect the mean spectrum and initialise guesses for the number of components, along with the amplitude, velocity, and FWHM of each component. These guesses are passed to \texttt{pyspeckit} via the \texttt{specfit} functionality, which returns the best fits to the specified components.

\begin{figure}[p]
    \centering

    \includegraphics[height=\figureHeight]{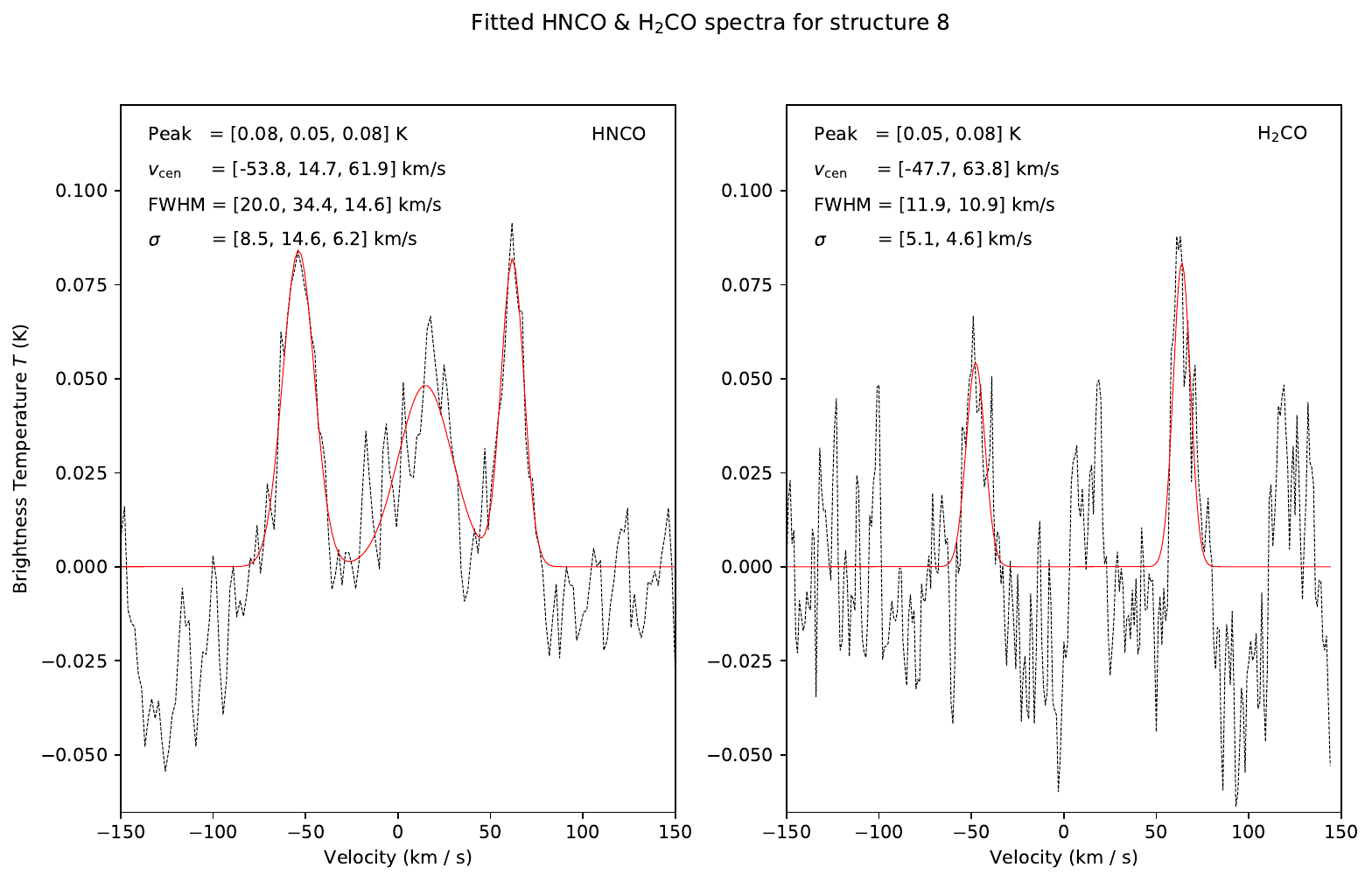}
    \label{fig:subfig1}

    \vspace{0cm}

    \includegraphics[height=\figureHeight]{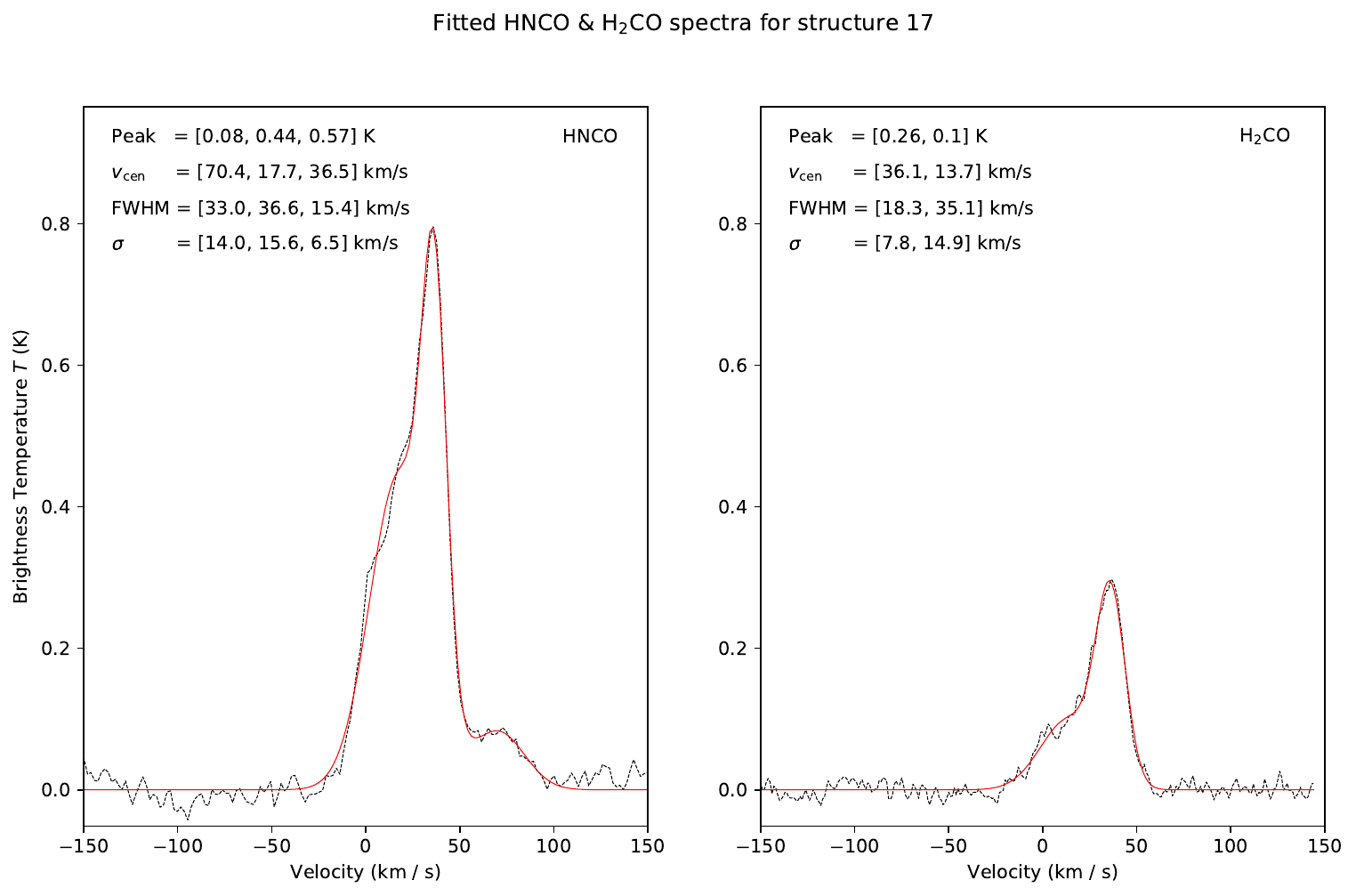}
    \label{fig:subfig2}

    \vspace{0cm}

    \includegraphics[height=\figureHeight]{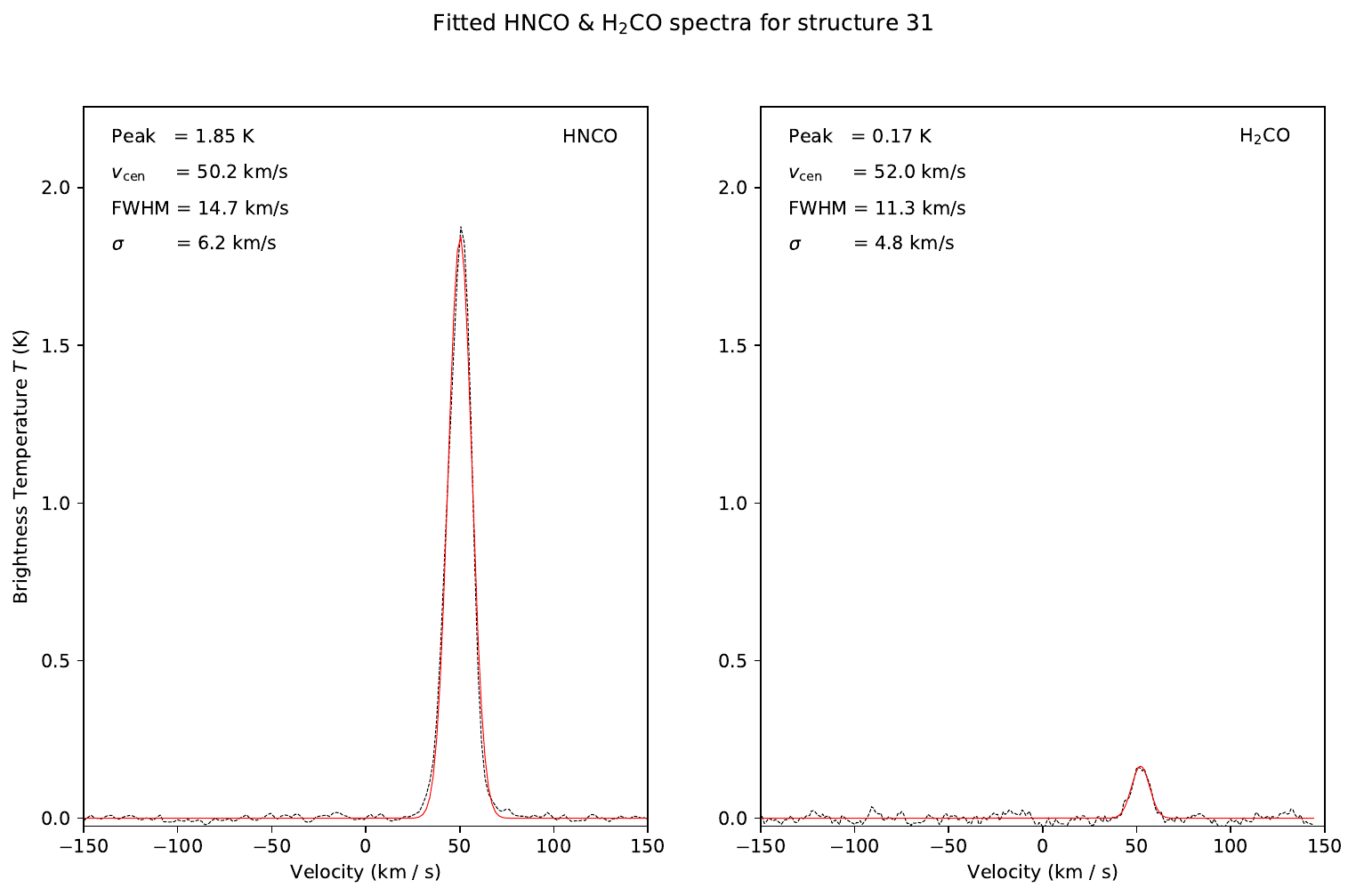}
    \label{fig:subfig3}

    \caption{Spatially averaged spectra and associated Gaussian fits for HNCO (left) and H$_{2}$CO (right), for sources 8 (top), 17 (centre), and 31 (bottom). The black dotted line shows the data, and the red solid line shows the resulting fit. Fit parameters (amplitude, centroid velocity, FWHM, and velocity dispersion) are shown in the legend in the upper left of each panel.}
    \label{fig:specfits}
\end{figure}

Figure \ref{fig:specfits} shows examples of the fits to the HNCO and H$_{2}$CO emission for three sources (8, 17, and 31). These sources are selected to demonstrate the range in spectral complexity in terms of varying numbers of components, blended components, and cases of poor signal to noise. This also supports the choice of HNCO as our main tracer, as it is almost always brighter than H$_{2}$CO, often revealing fainter velocity components that are not clearly seen otherwise.

\section{Absorption analysis: cases of near/far ambiguity}
\label{ap:abs}

In Section \ref{sec:fractional_abs} we highlighted that there are 8 cases for which we are unable to confidently determine whether a cloud is on the near of far side of the CMZ. The sources are 8a, 8c, 11a, 11b, 11c, 12, 15, and 24 (see Table \ref{tab:absorption}). Here we provide figures for each of these clouds to show the various absorption and emission features. The figures are equivalent to Figure \ref{fig:absorption_brick}.

Figure \ref{abs_8} shows source 8. The components 8a and 8c have systemic velocities of -54~\kms and 62~\kms, respectively. At these velocities we do see some absorption, but it is very low signal to noise, and does not show clear correlation in the PV slice. Combined with fractional absorption values that are slightly greater than 1, we conclude that their near/far assignment is uncertain.

Figure \ref{abs_11} shows source 11. The components 11a,b,c have systemic velocities of -11~\kms, 45~\kms, and 14~\kms, respectively. Though there is clear absorption towards the lower velocity components, there is very bright apparent emission at the velocity of the higher component. It is not clear whether this is real, as we don't necessarily expect bright emission in the line outside specific circumstances (e.g. H$_{2}$CO masers, none of which are seen at this location). This source is also very close to Sgr A* in projection, and has corresponding bright C-band continuum in the field. So while we acknowledge that there is potential absorption towards 11a and 11c, suggesting near side positions, we cannot confidently conclude this given the current data, and decide to mark this whole source as uncertain.

Figure \ref{abs_12} shows source 12. The systemic velocity is 86~\kms, at which we do see shallow absorption with a maximum depth of \app 6~K. However, this source is near to Sgr A* and the arched filaments in projection, which means that the C-band continuum and resulting fractional absorption values are large. So while the weak absorption suggests being on the near side, the fractional absorption suggest the opposite. We therefore do not assign a near/far placement for this source.

Figure \ref{abs_15} shows source 15. The systemic velocity is 52~\kms. Similarly to source 12, we do see absorption at this velocity, but due to the proximity to the arched filaments in projection, we also have bright C-band continuum, leading to large fractional absorption values. There is also apparent H$_{2}$CO (1$_{1,0}-1_{1,1}$) emission here, but again we do not know if it is real emission or an artifact. The apparent emission and absorption of H$_{2}$CO (1$_{1,0}-1_{1,1}$) are both strongly correlated in position-velocity, across different positions but the same velocity range. Given all of these uncertainties, we do not assign any near/far position for this source.

Finally, Figure \ref{abs_24} shows source 24. The systemic velocity is 53~\kms. While there is some weak absorption around this systemic velocity, and within the velocity dispersion of the source, it does not peak at 53~\kms. In fact, there is apparent emission at this velocity. This source is very close to the main Sgr B2 complex, both in position and velocity, and is likely part of the larger cloud complex which connects to the dust ridge. In this case, it could be argued that it is likely on the near side of the CMZ. However, given that the Sgr B2 complex is kinematically complicated, and the fact that we do not see deep absorption at the systemic velocity, we do not assign any near/far position for this source.

\foreach \x in {8,11,12,15,24}
{
\begin{figure*}
\includegraphics[width=\textwidth]{\x.pdf}
\caption{Overview of the emission and absorption towards source \textbf{\x}. \textit{Top left}: Cloud-averaged spectra of $^{13}$CO (dashed black line) and H$_{2}$CO (1$_{1,0}-1_{1,1}$) (solid black line). \textit{Top right}: PV slice of H$_{2}$CO (1$_{1,0}-1_{1,1}$) taken along the major axis of the cloud. \textit{Bottom left}: Integrated H$_{2}$CO (1$_{1,0}-1_{1,1}$) intensity. \textit{Bottom centre}: Integrated HNCO (4$_{0,4}$ - 3$_{0,3}$) intensity. \textit{Bottom right}: C-band (4.85~GHz) continuum.}
\expandafter\label\expandafter{abs_\x}
\end{figure*}
\clearpage
}

\bibliography{refs}{}
\bibliographystyle{aasjournal}
\end{document}